\RequirePackage{fix-cm}
\documentclass[smallextended]{svjour3}       
\smartqed  
\usepackage{graphicx}
\usepackage{amsmath}
\usepackage{amssymb}
\usepackage{physics}

\newcommand{\zerovec}{\ensuremath{\boldsymbol{0}}}

\usepackage{color}

\begin{document}

\title{Stationary measure induced by the eigenvalue problem of the one-dimensional 
Hadamard walk 
}
\subtitle{}

\titlerunning{Stationary measure induced by the eigenvalue problem}        

\author{Takashi Komatsu\footnote{T.~Komatsu\\
Department of Bioengineering School of Engineering, The University of Tokyo, Bunkyo, Tokyo, 113-8656, Japan\\
 \email{komatsutakashi@g.ecc.u-tokyo.ac.jp} (e-mail of the corresponding author)} \and Norio Konno}

\authorrunning{T.~Komatsu and N.~Konno} 

\institute{%
%
N.~Konno \at
Department of Applied Mathematics, Faculty of Engineering, Yokohama National University, 79-5 Tokiwadai, Hodogaya, Yokohama, 240-8501, Japan\\
              \email{konno-norio-bt@ynu.ac.jp}
}

\date{Received: date / Accepted: date}

\maketitle 

\begin{abstract}
In this paper, we consider the stationary measure of the Hadamard walk on the one-dimensional integer lattice. Here all the stationary measures given by solving the eigenvalue problem are completely determined via the transfer matrix method. Then these stationary measures can be divided into three classes, i.e., quadratic polynomial, bounded, and exponential types. In particular, we present an explicit necessary and sufficient condition for the bounded-type stationary measure to be periodic.  
\end{abstract}
\section{Introduction \label{intro}}
The notion of quantum walks was introduced by Aharonov et al. \cite{adz} as a quantum counterpart of the classical one-dimensional random walks. It is known that the long-time asymptotic behavior of the transition probability for quantum walks on the one-dimensional lattice is quite different from that of classical random walks \cite{ko1}. Recently, the quantum walk is intensively studied in various fields  \cite{mw,por}. Hadamard walk and Grover walk are the most fundamental objects in this field. For example, the Grover walk was applied to spatial search algorithms \cite{akr15}.

In this paper, we focus on a sequence of measures $\{\mu_{n}\}_{n\in\mathbb{Z}_{\geq}}$ induced by the unitary operator (time evolution operator) $U$ for quantum walks, where $\mathbb{Z}_{\geq}=\{0,1,2,\ldots\}$. Especially, one of the basic interests for quantum walks is to determine measures which do not depend on time $n\in\mathbb{Z}_{\geq}$, that is to say, our purpose is to obtain  measures satisfied with $\mu_{0}=\mu_{n}$ for $n\in\mathbb{Z}_{\geq}$. These measures are called the {\it stationary measure}. We can obtain stationary measures by using generalized eigenfunction \eqref{ef1}.  
\begin{eqnarray}\label{ef1}
Uf=e^{i\theta}f\qquad (f\in \ell^{\infty}(\mathbb{Z},\mathbb{C}^{2}),\ e^{i\theta}\in S^{1}).
\end{eqnarray}

Mainly, there are two motivations to study stationary measures. One is a relationship between scattering matrices and stationary measures. Recently, quantum walks have been intensively studied in terms of the spectral analysis \cite{hm,s1}. This point of view is based on the scattering theory of quantum mechanics like Schr\"{o}dinger equations. One of the most famous quantum effects is the quantum tunneling \cite{M1}. This effect shows that a quantum particle can tunnel through a barrier that it classically could not surmount. We can observe this phenomena to the quantum walks \cite{KKMS1,MMOS}. 
The scattering matrix is represented as the amplitudes of the reflected wave and the transmitted wave, and naturally appears in generalized eigenfunction for the time evolution operator \cite{hm}. On the other hand, stationary measures is derived from the generalized eigenfunction Eq. \eqref{ef1}. In terms of stationary measures, we can discuss quantum effects. Secondly, there is a relationship between spectrum $\sigma(U_{C})$ and stationary measures. The Hadamard walk is characterized the following way. 
\begin{itemize}
\item [$\bullet$] If the stationary measure induced by some $\lambda\in S^{1}$ is bounded type, $\lambda$ is a continuous spectrum. 
\item [$\bullet$] If the stationary measure induced by some $\lambda\in S^{1}$ is quadratic polynomial type, $\lambda$ lie in boundary of continuous spectrums. 
\item [$\bullet$] If the stationary measure induced by some $\lambda\in S^{1}$ is exponential type, $\lambda$ is in the resolvent set. 
\end{itemize}
Thus, from stationary measure's point of view, it is important to investigate spectrum of $U_{C}$.

The first result of stationary measures for quantum walks is given by Konno et al. \cite{kls}. The intensive study on stationary measures for quantum walks was reported and it is shown that there exists the uniform measure as stationary measure on regular graphs in Konno \cite{ko2}. That is to say, Konno proved that the set of uniform measures is contained the set of stationary measures. After that Konno and Takei \cite{kt} gave non-uniform stationary measures. In our previous work \cite{Kawai2017}, we investigated the stationary measures for the three-state quantum walks including the Fourier and Grover walks by solving the corresponding eigenvalue problem. Then we found the stationary measure with a periodicity. Recently, Komatsu and Konno \cite{Komatsu2017} obtained the stationary measure for quantum walks on the higher-dimensional integer lattice. 

Mathematically, it is important to consider the correspondence with classical random walks. For example, it is well known that the stationary measures $\mu$ for the classical random walks on the one dimensional integer lattice are given by
\[
\mu(x)=
\begin{cases}\vspace{0.3cm}
C>0& \cdots\ \mbox{uniform measure}\\
\left(\displaystyle\frac{q}{p}\right)^{x}&\cdots\ \mbox{exponential type}
\end{cases}\qquad(x\in\mathbb{Z}),
\]
where $p$, $q\in\mathbb{R}_{\geq}$ and $0<p<q$, $p+q=1$. Here, $\mathbb{R}_{\geq}$ is the set of the non negative real numbers. In this paper, we study stationary measures for quantum walks. Here, in order to clarify a one-to-one relationship between stationary measures of classical random walks and one of quantum walks, we introduce the corresponding singed measure of random walks in Type B of the following Table 1.\\

\vspace{0.2cm}
\begin{table}[h]
\begin{center}
\caption{Stationary measures}

\begin{tabular}{|c|c|c|} \hline

$ $ & \mbox{Random} \mbox{walk} & \mbox{Quantum}\ \mbox{walk}\\ \hline

Type A& \begin{tabular}{c}Uniform measure\\
$\mu(x)=c\in(0,\infty)$\end{tabular} & \begin{tabular}{c} Bounded type\\
\mbox{(Including uniform measures)}\end{tabular} \\ \hline

Type B&\begin{tabular}{c}First-degree polynomial\\
$\mu(x)=c_{1}x+c_{2}$\end{tabular} & \begin{tabular}{c}\mbox{Quadratic}\ \mbox{polynomial}\ type\\
$\mu(x)=c_{1}x^{2}+c_{2}x+c_{3}$\end{tabular}\\ \hline

Type C&\begin{tabular}{c}\mbox{Exponential}\ type\\
$\mu(x)=\left(\displaystyle\frac{q}{p}\right)^{x}$\end{tabular}&
\begin{tabular}{c}\mbox{Exponential}\ type\\
$\mu(x)=c^{x}$\end{tabular}\\ \hline

\end{tabular}

\end{center}

\end{table}
\noindent Note that random walks are $L^{1}$-norm and quantum walks are $L^{2}$-norm. On the other hand, in our previous work \cite{Komatsu2017}, we discussed the stationarity of the Grover walks on the $d$-dimensional integer lattice. In this case, we showed that the Grover walk has probability measures with the stationarity. More precisely, there exists the stationary measure with a finite support for the Grover walk. 
 


The purpose of this paper is to determine the set of the stationary measures induced by the eigenvalue problem for the Hadamard walk. Our method is based on the transfer matrices introduced by Kawai et al. \cite{Kawai2018}. The following results will be proved by applying propositions 
obtained in the subsequent section. 

We have the following two main results. 

\noindent{\bf{Result\ 1.}} $($Theorem $2$$\ \mbox{in}\ \mbox{Sect.}\ $5$)$ The set of the stationary measures induced by the eigenvalue problem for the Hadamard walk on $\mathbb{Z}$ is divided into three classes, where $\mathbb{Z}$ is the set of integers. One is the set of the measures with quadratic polynomial type. The second one is the set of the measures with bounded type. The last one is the set of the measures with exponential type. 
\vspace{0.3cm}

\noindent{\bf{Result\ 2.}} $($Theorem $3$$\ \mbox{in}\ \mbox{Sect.}\ $5$)$ An explicit necessary and sufficient condition for the bounded-type stationary measure to be periodic is given.

\vspace{0.3cm}
The rest of this paper is organized as follows. Section \ref{defqw} is devoted to the definition of the space-homogeneous quantum walk on the one-dimensional integer lattice. In Sect. \ref{method}, the transfer matrices given by Kawai et al. \cite{Kawai2018} to analyze stationary measures are defined and we collect some general facts from \cite{Kawai2018}. We discuss some aspects of the stationary measures induced by the general coin matrices in Sect. 4. More precisely, the set of the measures with a stationarity is decomposed into three classes. In Sect. \ref{result}, we give more detail formula by using symmetry of the Hadamard walk. In particular, we present an explicit necessary and sufficient condition for the stationary measure to be periodic. 
Conclusions are given in the last section.  

\section{Definition of the quantum walks on $\mathbb{Z}$}\label{defqw}
In this section, we give the definition of two-state quantum walk on $\mathbb{Z}$. A particle in the classical random walk moves at each step either one unit to the right with probability $p$ or one unit to the left with probability $q$, where $p+q=1$, $p$, $q>0$. On the other hand, the discrete-time quantum walk describes not only the motion of a particle but also the change of the states of a particle.

In the present paper, we consider the discrete-time quantum walk on $\mathbb{Z}$ defined by a unitary operator $U_{C}$ of the following form
\begin{equation}\label{u=sc}
U_{C}=SC,
\end{equation}
where the shift operator $S$ is given by
\[
S=\tau^{-1}\begin{bmatrix}
1&0\\
0&0\\
\end{bmatrix}+\tau \begin{bmatrix}
0&0\\
0&1\\
\end{bmatrix}.
\]
Here, the operator $\tau$ is defined by
\[
(\tau f)(x)=f(x-1)\qquad(f:\mathbb{Z}\longrightarrow\mathbb{C}^{2},\ x\in\mathbb{Z}),
\]
and $C$ is the following $2\times2$ unitary matrix
\begin{align}\label{coinm}
C=\begin{bmatrix}
c_{11}&c_{12}\\
c_{21}&c_{22}
\end{bmatrix}.
\end{align}
We call this unitary matrix the {\it coin matrix}. To consider the time evolution Eq. \eqref{u=sc}, decompose the matrix $C$ as
\[C=P+Q\]
with
\[P\equiv\begin{bmatrix}
1&0\\
0&0\\
\end{bmatrix}C=\begin{bmatrix}
c_{11}&c_{12}\\
0&0
\end{bmatrix},\quad 
Q\equiv\begin{bmatrix}
0&0\\
0&1\\
\end{bmatrix}C=\begin{bmatrix}
0&0\\
c_{21}&c_{22}\\
\end{bmatrix}
.
\]
We put $\Delta$ and $\widetilde{\Delta}$ as follows;
\begin{equation}\label{coin3}
\Delta=\det(A)=c_{11}c_{22}-c_{12}c_{21},\qquad \widetilde{\Delta}=c_{11}c_{22}+c_{12}c_{21}.
\end{equation}
The above Eq. \eqref{coin3} is utilized in Sect. $3$. Let $\mathbb{C}$ be the set of complex numbers. The state at time $n$ and location $x$ can be expressed by a two-dimensional vector:
\[\Psi_{n}(x)=\begin{bmatrix}\vspace{0.3cm}\Psi^{L}_{n}(x)\\ \Psi^{R}_{n}(x) \end{bmatrix}\in\mathbb{C}^{2}\quad (x\in\mathbb{Z},\ n\in\mathbb{Z}_{\geq}).\]
The time evolution of a quantum walk with a coin matrix $C$ is defined by the unitary operator $U_C$ in the following way:
\begin{equation}\label{timeevo}
\Psi_{n+1}(x)\equiv(U_C\Psi_{n})(x)=P\Psi_{n}(x+1)+Q\Psi_{n}(x-1).
\end{equation}
 This equation means that the particle moves at each step one unit to the right with matrix $Q$ or one unit to the left with matrix $P$. Let $\mathbb{R}_{\geq}=[0,\infty)$. For time $n\in\mathbb{Z}_{\geq}$, we define the measure $\mu_{n}\ :\ \mathbb{Z}\longrightarrow \mathbb{R}_{\geq}$ by
$$\mu_n(x)=\|\Psi_n(x)\|_{\mathbb{C}^2}^2,$$
where $\|\cdot\|_{\mathbb{C}^2}$ denotes the standard norm on $\mathbb{C}^2$. Let $\mathcal{M}(U_C)$ be the set of measures on $\mathbb{Z}$, where $U_C$ is a unitary operator given by Eq. \eqref{timeevo}.

Let $\mbox{Map}(\mathbb{Z},\mathbb{C}^{2})$ be the set of the functions from $\mathbb{Z}$ to $\mathbb{C}^{2}$. Now we define an operator $\phi$
\[
\begin{array}{ccc}
\phi: \mbox{Map}(\mathbb{Z},\mathbb{C}^{2})\setminus\{\zerovec\} & {\longrightarrow} & \mathcal{M}(U_{C})\\
\hspace{0.45cm} 

 \rotatebox{90}{$\in$} & & \rotatebox{90}{$\in$} \\
 \hspace{0.45cm} 
 
\Psi & \longmapsto & \mu
\end{array}
\]
 such that for $x\in\mathbb{Z}$ and $\Psi\ne\zerovec\in\mbox{Map}(\mathbb{Z},\mathbb{C}^{2})$, 
$$\phi(\Psi)(x)=|\Psi^{L}(x)|^2+|\Psi^{R}(x)|^2,\qquad\left(\Psi(x)=\begin{bmatrix}\Psi^{L}(x)\\ \Psi^{R}(x)\\ \end{bmatrix}\right).$$
From the above definition, we denote $\mu:=\phi(\Psi)\in\mathcal{M}(U_{C})$. 
\section{Stationary measure and Transfer matrix}\label{method}
\subsection{Definition of stationary measure for quantum walk}
\noindent
In this section, we  discuss a sequence of measures $\{\mu_{n}\}_{n\in\mathbb{Z}_{\geq}}$ induced by the unitary operator $U_{C}$ for quantum walks. Especially, we focus on the a sequence of measures $\{\mu_{n}\}_{n\in\mathbb{Z}_{\geq}}$ with a stationarity, namely
\[
\mu_{0}=\mu_{1}=\cdots=\mu_{n}=\cdots\qquad(n\in\mathbb{Z}_{\geq}).
\] 
In other words, the measure with a stationarity is a non-negative real-valued function on $\mathbb{Z}$ that does not depend on the time $n\in\mathbb{Z}_{\geq}$. We put the set of the stationary measures $\mathcal{M}_s(U_C)$ as  
\begin{equation*}
\begin{split}
&\mathcal{M}_s(U_C)=\Big\{\mu\in \mathcal{M}(U_C): \mbox{there}\ \mbox{exists}\ \Psi_0\in\mbox{Map}(\mathbb{Z},\mathbb{C}^{2})\ \mbox{such}\ \mbox{that}\\
&\hspace{7.5cm}\mu=\phi(U_C^n\Psi_0)\ (n\in\mathbb{Z}_{\geq})\Big\}.
\end{split}
\end{equation*}
We call this measure $\mu\in\mathcal{M}_s(U_C)$ the stationary measure for the quantum walk defined by the unitary operator $U_C$. If $\mu\in\mathcal{M}_s(U_C)$, then $\mu_n=\mu$ for $n\in\mathbb{Z}_{\geq}$, where $\mu_n$ is the measure of quantum walk given by $U_C$ at time $n$. 

In general, if unitary operators $U_{C_{1}}$ and $U_{C_{2}}$ are different, the sets of stationary measures $\mathcal{M}_s(U_{C_{1}})$ and $\mathcal{M}_s(U_{C_{2}})$ are different. For example, if we take the unitary operators $U_{C_{1}}$ and $U_{C_{2}}$ corresponding to the following coin matrices $C_1$ and $C_2$ respectively:
\begin{align*}
C_1=
\begin{bmatrix}
1&0\\
0&1
\end{bmatrix},\qquad
C_2=\frac{1}{\sqrt{2}}
\begin{bmatrix}
1&1\\
1&-1
\end{bmatrix},
\end{align*}
then we have
\begin{align*}
\mathcal{M}_s(U_{C_{1}})=\mathcal{M}_{unif}(U_{C_{1}}),\ \ \ 
\mathcal{M}_s(U_{C_{2}})\supsetneq\mathcal{M}_{unif}(U_{C_{2}}).
\end{align*}
The above results are given in Konno and Takei \cite{kt}. Here $\mathcal{M}_{unif}(U_{C})$ is the set of the uniform measures defined by
\begin{equation}\label{unif1}
\begin{split}
&\mathcal{M}_{unif}(U_{C})=\Big\{\mu_{c}\in\mathcal{M}(U_{C}):\mbox{there}\ \mbox{exists}\ c>0\\
&\hspace{5.5cm}  \mbox{such}\ \mbox{that}\ \mu_{c}(x)=c\ (x\in\mathbb{Z})\Big\}.
\end{split}
\end{equation}
\subsection{Transfer matrix induced by the eigenvalue problem}
We define the transfer matrices to analyze stationary measures for quantum walks in this subsection. A method based on transfer matrices is one of the common approaches, for example, Ahlbrecht et al. \cite{Ahl2011}, Bourget et al. \cite{Bou2003} and Kawai et al. \cite{Kawai2018}. In this paper, we apply this method to two-state space-homogeneous quantum walks to obtain the stationary measures.

Let $S^{1}\subset\mathbb{C}$ be the following unit circle in complex plane.
\[
S^{1}=\left\{z\in\mathbb{C}:|z|=1\right\}.
\]
Now we consider the following eigenvalue problem of the quantum walk determined by $U_C$:
\begin{equation}\label{eig.pro}
U_C\Psi=\lambda\Psi\quad(\lambda\in S^{1}).
\end{equation}
Then we see that $U_C\Psi=\lambda\Psi$ is equivalent to the following relations:
\begin{equation}\label{equation1}
 \begin{cases}
      \lambda\Psi^L(x)=c_{11}\Psi^L(x+1)+c_{12}\Psi^{R}(x+1) ,                                                                                
          \\
     \lambda\Psi^R(x)=c_{21}\Psi^L(x-1)+c_{22}\Psi^{R}(x-1) .                                                                                   
      \end{cases}
\end{equation}
Suppose that $c_{11}\ne0$. Remark that $c_{11}\ne0$ gives $c_{22}\ne0$. From above Eq. \eqref{equation1}, we get
\begin{equation}\label{trans1}
\begin{cases}
\bullet\begin{bmatrix}\displaystyle \Psi^{L}(x)\\ \Psi^{R}(x) \end{bmatrix}=
\begin{bmatrix}\vspace{0.4cm}
\displaystyle\frac{\lambda^2-c_{12}c_{21}}{c_{11}\lambda}&\displaystyle-\frac{c_{12}c_{22}}{c_{11}\lambda}\\
\displaystyle\frac{c_{21}}{\lambda}&\displaystyle\frac{c_{22}}{\lambda}
\end{bmatrix}\vspace{0.4cm}
\begin{bmatrix}\Psi^{L}(x-1)\\ \Psi^{R}(x-1) \end{bmatrix},
          \\
     \bullet\begin{bmatrix}\displaystyle \Psi^{L}(x)\\  \Psi^{R}(x) \end{bmatrix}=
\begin{bmatrix}\vspace{0.4cm}
\displaystyle \frac{c_{11}}{\lambda}&\displaystyle\frac{c_{12}}{\lambda}
\\
\displaystyle-\frac{c_{11}c_{21}}{\lambda}&\displaystyle \frac{\lambda^2-c_{12}c_{21}}{c_{22}\lambda}
\end{bmatrix}
\begin{bmatrix}\Psi^{L}(x+1)\\ \Psi^{R}(x+1) \end{bmatrix} .                                                                                 
      \end{cases}
      \end{equation}
Hence we put the following matrices $T^{+}_{\lambda}(C)$, $T^{-}_{\lambda}(C)$ as
\begin{eqnarray}\label{trans2}
T^{+}_{\lambda}(C)=\begin{bmatrix}\vspace{0.4cm}
\displaystyle\frac{\lambda^2-c_{12}c_{21}}{c_{11}\lambda}&\displaystyle-\frac{c_{12}c_{22}}{c_{11}\lambda}\\
\displaystyle\frac{c_{21}}{\lambda}&\displaystyle\frac{c_{22}}{\lambda}
\end{bmatrix}\vspace{0.4cm},\quad 
T^{-}_{\lambda}(C)=\begin{bmatrix}\vspace{0.4cm}
\displaystyle \frac{c_{11}}{\lambda}&\displaystyle\frac{c_{12}}{\lambda}
\\
\displaystyle-\frac{c_{11}c_{21}}{c_{22}\lambda}&\displaystyle \frac{\lambda^2-c_{12}c_{21}}{c_{22}\lambda}
\end{bmatrix}.
\end{eqnarray}
We call these matrices the transfer matrices. These matrices have the following relation:
\[
T^{+}_{\lambda}(C)\ T^{-}_{\lambda}(C)=T^{-}_{\lambda}(C)\ T^{+}_{\lambda}(C)=I,
\]
where $I$ is the identity matrix. It should be remarked that the transfer matrices defined by Eq. \eqref{trans2} are not always unitary. If $T^{+}_{\lambda}(C)$ is a unitary matrix, the stationary measure induced by the transfer matrices is a uniform measure, because a unitary matrix preserves the norm. However, the converse is not true. In Sect. \ref{result}, this counterexample is given by the Hadamard walk. 

 We write $\Psi(0)$ $(\Psi\in \mbox{Map}(\mathbb{Z},\mathbb{C}^{2})\setminus\{\zerovec\})$ as 
\begin{equation}\label{initial1}
\Psi(0)=\begin{bmatrix}\Psi^{L}(0)\\ \Psi^{R}(0)\\ \end{bmatrix}
=\begin{bmatrix}\varphi_{1}\\ \varphi_{2}\\ \end{bmatrix}
\quad\left(\varphi_{1},\varphi_{2}\in\mathbb{C}\right).
\end{equation}
From Eqs. \eqref{trans1} and \eqref{initial1}, we get
\begin{equation}\label{trans3}
\begin{split}
&\Psi^{L}(1)=\frac{\displaystyle\varphi_{1}\lambda^{2}-c_{12}(c_{21}\varphi_{1}+c_{22}\varphi_{2})}{c_{11}\lambda},\qquad
\Psi^{R}(1)=\frac{c_{21}\varphi_{1}+c_{22}\varphi_{2}}{\lambda},\\
&\Psi^{L}(-1)=\frac{c_{11}\varphi_{1}+c_{12}\varphi_{2}}{\lambda},\qquad
\Psi^{R}(-1)=\frac{\displaystyle\varphi_{2}\lambda^{2}-c_{21}(c_{11}\varphi_{1}+c_{12}\varphi_{2})}{c_{22}\lambda}.
\end{split}
\end{equation}
The above Eq. \eqref{trans3} will be used in Sect. $5$. 

The purpose of this paper is to find stationary measures for our two-state quantum walks by using Eq. \eqref{trans2}. Here we define a subset $\mathcal{M}_{s}^{(\lambda)}(U_{C})$ of $\mathcal{M}_{s}(U_{C})$ as
\[
\mathcal{M}_{s}^{(\lambda)}(U_{C})=\left\{\mu\in\mathcal{M}_{s}(U_{C}):\mu=\phi(\Psi)\ \mbox{such\ that} \ U_{C}\Psi=\lambda\Psi\right\}\qquad(\lambda\in S^{1}).
\]
We put the set of collection stationary measure induced by the eigenvalue problem as
\[
\widetilde{\mathcal{M}_{s}(U_{C})}=\bigcup_{\lambda\in S^{1}}\mathcal{M}_{s}^{(\lambda)}(U_{C}).
\]
For $\Psi\in\mbox{Map}(\mathbb{Z},\mathbb{C}^{2})\setminus\{\zerovec\}$ with Eq. \eqref{eig.pro}, we note that 
\begin{equation}\label{eig.pro2}
\phi(\Psi)\in\mathcal{M}_{s}(U_{C}).
\end{equation}
\subsection{Previous study}
In this subsection, we give some subsets of $\mathcal{M}_s(U_C)$ and briefly explain the previous study on stationary measures for quantum walks. 

Now, we prepare some classes of the set of the stationary measures to explain our results. First one is the set of the measures with exponential type $\mathcal{M}_{s,exp}(U_{C})$, i.e., 
\begin{equation*}
\begin{split}
&\mathcal{M}_{s,exp}(U_{C})=\Big\{\mu\in\mathcal{M}_{s}(U_{C}):\mbox{there\ exist}\ c_{+},\ c_{-}>0\ (c_{+}, c_{-}\ne1)\\
&\hspace{2.5cm}\mbox{such\ that}\ 0<\lim_{x\to+\infty}\frac{\mu(x)}{c_{+}^x}<+\infty,\quad 0<\lim_{x\to-\infty}\frac{\mu(x)}{c_{-}^x}<+\infty\Big\}.
\end{split}
\end{equation*}
We put the set $\widetilde{\mathcal{M}_{s,exp}(U_{C})}$ as
\[
\widetilde{\mathcal{M}_{s,exp}(U_{C})}=\mathcal{M}_{s,exp}(U_{C})\cap\widetilde{\mathcal{M}_{s}(U_{C})}.
\]
Second one is the set of the measures with quadratic polynomial type $\mathcal{M}_{s,qp}(U_{C})$, i.e.,
\begin{equation*}
\begin{split}
&\mathcal{M}_{s,qp}(U_{C})=\Big\{\mu\in\mathcal{M}_{s}(U_{C}):0<\lim_{x\to\pm\infty}\frac{\mu(x)}{|x|^2}<+\infty\Big\}.
\end{split}
\end{equation*}
We put the set $\widetilde{\mathcal{M}_{s,qp}(U_{C})}$ as
\[
\widetilde{\mathcal{M}_{s,qp}(U_{C})}=\mathcal{M}_{s,qp}(U_{C})\cap\widetilde{\mathcal{M}_{s}(U_{C})}.
\]
The last one is the set of the uniform measures given by Eq. \eqref{unif1}. The uniform measure is a positive real-valued constant function on $\mathbb{Z}$. In other words, we can regard a uniform measure as a measure with period $1$. Therefore, we define the subset $\mathcal{M}_{s,period}^{(m)}(U_{C})$ as
\[
\mathcal{M}_{s,period}^{(m)}(U_{C})=\{\mu\in\mathcal{M}_{s}(U_{C}):\mu(x+m)=\mu(x)\ (x\in\mathbb{Z})\}.
\]
Here, $m\in\mathbb{N}$. It is remarked that
\[
\mathcal{M}_{s,period}^{(1)}(U_{C})=\mathcal{M}_{unif}(U_{C})\cap\mathcal{M}_{s}(U_{C}).
\]
Moreover, we set the subset $\mathcal{M}_{s,bdd}(U_{C})$ of $\mathcal{M}_{s}(U_{C})$ as
\begin{equation*}
\begin{split}
&\mathcal{M}_{s,bdd}(U_{C})=\{\mu\in\mathcal{M}_{s}(U_{C}):\mbox{there\ exists}\ M>0\\
&\hspace{6.5cm} \mbox{such\ that}\ \mu(x)\leq M (x\in\mathbb{Z})\}.
\end{split}
\end{equation*}
We put the set $\widetilde{\mathcal{M}_{s,bdd}(U_{C})}$ as
\[
\widetilde{\mathcal{M}_{s,bdd}(U_{C})}=\mathcal{M}_{s,bdd}(U_{C})\cap\widetilde{\mathcal{M}_{s}(U_{C})}.
\]
We briefly review the result of our previous work in \cite{Kawai2018}.  
\begin{theorem}[Corollary 3.4 in \cite{Kawai2018} ]\label{kawaithm}
Let $\lambda\in S^{1}$ be an eigenvalue satisfied with Eq. \eqref{eig.pro}. 
We put the function $\Psi\in\mbox{Map}(\mathbb{Z},\mathbb{C}^{2})\setminus\{\zerovec\}$ and write 
\[
\Psi(x)=\begin{bmatrix}\Psi^{L}(x)\\ \Psi^{R}(x)\\ \end{bmatrix}\qquad(x\in\mathbb{Z}).
\]
For a coin matrix $C$ defined by Eq. \eqref{coinm} with $c_{11}\neq0$, a solution of the eigenvalue problem induced by Eq. \eqref{eig.pro} is given in the following.
\begin{description}
\item[$(i)$] For case of $\lambda^2 = c_{11}c_{22}+c_{12}c_{21}\pm2\sqrt{c_{11}c_{12}c_{21}c_{22}}$, we get
\begin{align*}
&\!\!\!\!\!\Psi(x)=\begin{bmatrix}
\Psi^{L}(x)\\
\Psi^{R}(x)
\end{bmatrix}\\
&=
\begin{cases}
\bigg(\dfrac{\lambda^2+\Delta}{2c_{11}\lambda}\bigg)^x \dfrac{1}{\lambda^2+\Delta}
\begin{bmatrix}
\varphi_{1}(1+x)\lambda^2-(\varphi_{1}\widetilde{\Delta}+2c_{12}c_{22}\varphi_{2})x+\varphi_{1}\Delta\vspace{3mm}\\
\varphi_{2}(1-x)\lambda^2+(\varphi_{2}\widetilde{\Delta}+2c_{11}c_{21}\varphi_{1})x+\varphi_{2}\Delta
\end{bmatrix}&(x\geq1),\\\\
\bigg(\dfrac{\lambda^2+\Delta}{2c_{22}\lambda}\bigg)^{-x} \dfrac{1}{\lambda^2+\Delta}
\begin{bmatrix}
\varphi_{1}(1+x)\lambda^2-(\varphi_{1}\widetilde{\Delta}+2c_{12}c_{22}\varphi_{2})x+\varphi_{1}\Delta\vspace{3mm}\\
\varphi_{2}(1-x)\lambda^2+(\varphi_{2}\widetilde{\Delta}+2c_{11}c_{21}\varphi_{1})x+\varphi_{2}\Delta
\end{bmatrix}&(x\leq-1).
\end{cases}
\end{align*}\\
\item[$(ii)$] For case of $\lambda^2 \neq c_{11}c_{22}+c_{12}c_{21}\pm2\sqrt{c_{11}c_{12}c_{21}c_{22}}$, we get
\begin{align*}
&\!\!\!\!\!
\Psi(x)=\begin{bmatrix}
\Psi^{L}(x)\\
\Psi^{R}(x)
\end{bmatrix}\\
&=
\begin{cases}
\dfrac{1}{\Lambda_{+} -\Lambda_{-}}
\begin{bmatrix}
\Lambda^{x}_{+} (\Psi^{L}(1)-\Lambda_{-}\varphi_{1})-
\Lambda^{x}_{-}(\Psi^{L}(1)-\Lambda_{+}\varphi_{1})\vspace{3mm}\\
\Lambda^{x}_{+} (\Psi^{R}(1)-\Lambda_{-}\varphi_{2})-
\Lambda^{x}_{-}(\Psi^{R}(1)-\Lambda_{+}\varphi_{2})
\end{bmatrix}&(x\geq1),\\\\
\dfrac{1}{\Gamma_{+} -\Gamma_{-}}
\begin{bmatrix}
\Gamma^{-x}_{+} (\Psi^{L}(-1)-\Gamma_{-}\varphi_{1})-
\Gamma^{-x}_{-}(\Psi^{L}(-1)-\Gamma_{+}\varphi_{1})\vspace{3mm}\\
\Gamma^{-x}_{+} (\Psi^{R}(-1)-\Gamma_{-}\varphi_{2})-
\Gamma^{-x}_{-}(\Psi^{R}(-1)-\Gamma_{+}\varphi_{2})
\end{bmatrix}&(x\leq-1).
\end{cases}
\end{align*}
\end{description}
Here, we denote that $\Lambda_{\pm}$ and $\Gamma_{\pm}$ are expressed by
\begin{equation}\label{lam123}
\Lambda_{\pm}=\dfrac{h(\lambda)\pm\sqrt{h(\lambda)^2-4\lambda^{2}c_{11}c_{22}}}{2c_{11}\lambda},\qquad
\Gamma_{\pm}=\dfrac{h(\lambda)\pm\sqrt{h(\lambda)^2-4\lambda^{2}c_{11}c_{22}}}{2c_{22}\lambda},
\end{equation}
where $h(\lambda)$ is defined by $h(\lambda)=\lambda^{2}+\Delta$. Furthermore, the definitions of $\Delta$, $\widetilde{\Delta}$, $\Psi^{L}(\pm1)$ and $\Psi^{R}(\pm1)$ are given in Eqs. \eqref{coin3} and \eqref{trans3}.
\end{theorem}
By using Eq. \eqref{eig.pro2}, we have the following result.
\begin{corollary}
For $\Psi\ne\zerovec\in\mbox{Map}(\mathbb{Z},\mathbb{C}^{2})$ given by Theorem \ref{kawaithm}, we obtain
\[
\phi(\Psi)\in\mathcal{M}_{s}(U_{C}).
\]
\end{corollary}
\section{Stationary measures to the general coin matrices}
\noindent In this section, we state the properties of stationary measures induced by the general coin matrices $C=(c_{ij})$ with $c_{11}\ne0$. From  Eq. \eqref{equation1}, we obtain the following equation.
\begin{equation}\label{poly1}
\lambda\frac{c_{11}}{c_{12}}\Psi^{j}(x+2)+\left(c_{21}-\frac{c_{11}c_{22}}{c_{12}}-\frac{\lambda^{2}}{c_{12}}\right)\Psi^{j}(x+1)+\lambda\frac{c_{22}}{c_{12}}\Psi^{j}(x)=0\qquad(j=L,\ R).
\end{equation}
We consider the characteristic polynomial induced by Eq. \eqref{poly1}.

\begin{equation}\label{cheq}
x^{2}+lx+\frac{c_{22}}{c_{11}}=0,
\end{equation}
where $l$ is given by
\[
l=-\frac{1}{c_{11}}\left(\lambda+\frac{\Delta}{\lambda}\right),\qquad \left|\frac{c_{22}}{c_{11}}\right|=1.
\]
Let $\Lambda_{+}$ and $\Lambda_{-}$ be solutions for a characteristic polynomial defined by Eq. \eqref{cheq}. Then, the solutions $\Lambda_{+}$ and $\Lambda_{-}$ become any of the following Type 1, Type 2, and Type 3.
\begin{itemize}
\item [$\bullet$]  Type $1$\ $:$\ $|\Lambda_{+}|=|\Lambda_{-}|$=1, $\Lambda_{+}=\Lambda_{-}$ 
\item [$\bullet$] Type $2$\ $:$\  $|\Lambda_{+}|=|\Lambda_{-}|$=1, $\Lambda_{+}\ne\Lambda_{-}$. 
\item [$\bullet$] Type $3$\ $:$\  $|\Lambda_{+}|>1>|\Lambda_{-}|>0$ or $|\Lambda_{-}|>1>|\Lambda_{+}|>0$. 
\end{itemize}
From Eq. \eqref{lam123}, we put the subsets $K_{1}$, $K_{2}$ $\subset[0,2\pi)$ as
\[
K_{1}=\biggl\{\theta:|\Lambda_{+}|=|\Lambda_{-}|=1, \Lambda_{+}=\Lambda_{-}\biggl\},\  K_{2}=\biggl\{\theta:|\Lambda_{+}|=|\Lambda_{-}|=1, \Lambda_{+}\ne\Lambda_{-}\biggl\}.
\]
Let $K_{3}$ be 
\[ 
K_{3}=\biggl\{\theta\in[0,2\pi):|\Lambda_{+}|>1>|\Lambda_{-}|>0\ \rm{or}\ |\Lambda_{-}|>1>|\Lambda_{+}|>0\biggl\}.
\]
Furthermore we set the subsets  $\widetilde{K_{j}}\subset S^{1}$ as
\[
\widetilde{K_{j}}=\{e^{i\theta}: \theta\in K_{j}\},\qquad (j=1,2,3).
\]
\begin{proposition}\label{geneco}
Let $C=(c_{ij})_{i,j=1,2}$ be a unitary matrix given by Eq. \eqref{coinm} with $c_{11}\ne0$. Then stationary measures induced by the quantum walk $U_{C}$ have the following properties.
\begin{itemize}
\item [$(1)$] If we take $\lambda\in\widetilde{K_{1}}$, it holds that $\phi(\Psi)\in \mathcal{M}_{s,qp}(U_{C})$ for some initial state $\varphi$.
\item [$(2)$] Suppose that $\lambda\in S^{1}\setminus\widetilde{K_{1}}$ . Then we obtain the followings. 
\begin{itemize}
\item [$(a)$] Suppose that $|\varphi_{1}|$, $|\varphi_{2}|<\infty$. For $\lambda\in\widetilde{K_{2}}$, the stationary measures $\phi(\Psi)$ induced by the function $\Psi\in\mbox{Map}(\mathbb{Z},\mathbb{C}^{2})$ in Theorem \ref{kawaithm} $(ii)$ have the measures with bounded type. That is to say,
\[
\phi(\Psi)\in\mathcal{M}_{s,bdd}(U_{C}).
\]
\item [$(b)$] For $\lambda\in\widetilde{K_{3}}$, the stationary measures $\phi(\Psi)$ induced by the function $\Psi\in{\rm{\mbox{Map}}}(\mathbb{Z},\mathbb{C}^{2})$ in Theorem \ref{kawaithm} $(ii)$ have the measures with exponential type. That is to say,
\[
\phi(\Psi)\in\mathcal{M}_{s,exp}(U_{C}).
\]
\end{itemize}
\end{itemize}
\end{proposition}
{\bf{Proof.}} We show the statement $(1)$. Suppose that $\lambda\in S^{1}$ satisfied with $\lambda^2 = c_{11}c_{22}+c_{12}c_{21}\pm2\sqrt{c_{11}c_{12}c_{21}c_{22}}$. Then we get 
\[
\lambda^{2}+\Delta=2\Delta\left(|c_{11}|^{2}\pm i|c_{11}||c_{12}|\right).
\]
From $|\lambda^{2}+\Delta|^{2}=4|c_{11}|^{2}$ and $|\lambda^{2}+\Delta|^{2}=4|c_{22}|^{2}$, 
we have 
\begin{equation}\label{niji1}
\bigg|\dfrac{\lambda^2+\Delta}{2c_{11}\lambda}\bigg|^{2x}
=\left(\frac{4|c_{11}|^{2}}{4|c_{11}|^{2}}\right)^{x}=1,\qquad \bigg|\dfrac{\lambda^2+\Delta}{2c_{22}\lambda}\bigg|^{-2x}=\left(\frac{4|c_{22}|^{2}}{4|c_{22}|^{2}}\right)^{-x}=1.
\end{equation}
If we take an appropriate condition $\varphi$, the stationary measure $\phi(\Psi)$ is the following formula \eqref{niji2}. From Theorem \ref{kawaithm} and Eq. \eqref{niji1}, there exist non-zero constants $A_{i}(\lambda,C,\varphi)$ $(i=1,2,3)$ which do not depend on the parameter $x$ such that 
\begin{equation}\label{niji2}
(\phi(\Psi))(x)=A_{1}(\lambda,C,\varphi)x^{2}+A_{2}(\lambda,C,\varphi)x+A_{3}(\lambda,C,\varphi).
\end{equation}
Hence, we have
\[
\phi(\Psi)\in \mathcal{M}_{s,qp}(U_{C}).
\]
Next, we show the statement $(a)$. For $\theta\in K_{2}$, note that 
\[
|\Lambda_{+}|=1,\qquad |\Lambda_{-}|=1,\qquad \Lambda_{+}\ne\Lambda_{-}.
\]
Therefore, there exists $\xi\in(0,2\pi)$ such that
\begin{equation}\label{bdd123}
\Lambda_{+}\cdot\overline{\Lambda_{-}}=e^{i\xi}.
\end{equation}
From Theorem \ref{kawaithm} and Eq. \eqref{bdd123}, the stationary measure $\phi(\Psi)$ is the following formula \eqref{bdd1234}.
\begin{equation}\label{bdd1234}
(\phi(\Psi))(x)=A_{4}(\lambda,C,\varphi)+A_{5}(\lambda,C,\varphi)\Re\left((\Lambda_{+}\cdot\overline{\Lambda_{-}})^{x}A_{6}(\lambda,C,\varphi)\right).
\end{equation}
Here, $A_{i}(\lambda,C,\varphi)<\infty$ $(i=4,5,6)$ do not depend on the parameter $x$. Since $|\varphi_{1}|$, $|\varphi_{2}|<\infty$, we obtain 
\[
\phi(\Psi)\in \mathcal{M}_{s,bdd}(U_{C}).
\]
Finally, we show the statement $(b)$. For $\theta\in K_{3}$, it holds 
\begin{equation}\label{exp1}
|\Lambda_{+}|>1>|\Lambda_{-}|>0\qquad \Lambda_{+}\cdot\overline{\Lambda_{-}}=1.
\end{equation}
From Theorem \ref{kawaithm} and Eq. \eqref{exp1}, the stationary measure $\phi(\Psi)$ is the following formula \eqref{exp2}.
\begin{equation}\label{exp2}
(\phi(\Psi))(x)=A_{7}(\lambda,C,\varphi)(|\Lambda_{+}|^{2})^{x}+A_{8}(\lambda,C,\varphi)(|\Lambda_{-}|^{2})^{x}+A_{9}(\lambda,C,\varphi),
\end{equation}
where $A_{i}(\lambda,C,\varphi)$ $(i=7,8,9)$ do not depend on the parameter $x$. Therefore, we have
\[
\phi(\Psi)\in \mathcal{M}_{s,exp}(U_{C}).
\]
\hspace{\fill} $\square$
\vspace{0.3cm}

From Proposition \ref{geneco}, stationary measures induced by a quantum walk $U_{C}$ with a general coin matrix $C$ were divided three classes. In the next discussion, we present more detail formula by using symmetry of the Hadamard walk given by 
\[
H=\frac{1}{\sqrt{2}}
\begin{bmatrix}
1&1\\
1&-1
\end{bmatrix}.
\]
This matrix $H$ is called the Hadamard matrix. Furthermore, we give an explicit necessary and sufficient condition for the bounded-type stationary measure to be periodic.
\section{Results}\label{result}
In this section, we consider some aspects of stationary measures. More precisely, when we take the Hadamard coin as a coin matrix $C$, the set of the measures with a stationarity is decomposed into three classes. First one is the set of the measures with quadratic polynomial type. This part of our results is mentioned by Konno and Takei \cite{kt}. The second one is  the set of the measures with bounded type. Especially, we present an explicit necessary and sufficient condition for the bounded-type stationary measure to be periodic. The last one is the set of the measures with exponential type. The second and last sets are obtained in our paper for the first time.
The purpose of this section is to prove the following theorem.  
\begin{theorem}We consider the stationary measures induced by the eigenvalue problem for the Hadamard walk on $\mathbb{Z}$. Then, we have
\[
\widetilde{\mathcal{M}_{s}(U_{H})}=\widetilde{\mathcal{M}_{s,qp}(U_{H})}\cup\widetilde{\mathcal{M}_{s,bdd}(U_{H})}\cup\widetilde{\mathcal{M}_{s,exp}(U_{H})}. 
\] 
Here these symbols $\widetilde{\mathcal{M}_{s,qp}(U_{H})}$, $\widetilde{\mathcal{M}_{s,bdd}(U_{H})}$, and $\widetilde{\mathcal{M}_{s,exp}(U_{H})}$ are defined in Sect. 3.3. 
\end{theorem}
From now on, we prepare some lemmas and propositions. 
\subsection{Results of Type $1$, $2$, and $3$} 
From now on, we treat the following orthogonal matrix $O(\zeta)$ as a unitary matrix $C$
\begin{equation*}
O(\zeta)=
\begin{bmatrix}
c&s\\
s&-c
\end{bmatrix}\qquad(c,s\ne0),
\end{equation*}
with $c=\cos\zeta$ and $s=\sin\zeta$. Note that the quantum walk determined by $O(\pi/4)$ becomes
the Hadamard walk.
\begin{lemma}\label{resultlem}
Let $\lambda\in S^{1}$ be an eigenvalue in Eq. \eqref{eig.pro}. The solutions of the equation  $\lambda^2 = -c^{2}+s^{2}\pm2i\sqrt{c^{2}s^{2}}$ are given by 
\[
\lambda_{1}=e^{i\frac{\eta}{2}},\qquad \lambda_{2}=e^{i(\pi-\frac{\eta}{2})},\qquad \lambda_{3}=e^{i(\pi+\frac{\eta}{2})},\qquad \lambda_{4}=e^{i(2\pi-\frac{\eta}{2})}\qquad(\eta\in(0,\pi)),
\]
where 
\[
\cos\eta=-(c^{2}-s^{2}),\qquad \sin\eta=2cs.
\] 
\end{lemma}
Especially, we consider the Hadamard walk corresponding to the following orthogonal matrix 
\[
O(\pi/4)\equiv H=\frac{1}{\sqrt{2}}
\begin{bmatrix}
1&1\\
1&-1
\end{bmatrix}.
\]
From Lemma \ref{resultlem}, we have 
\begin{equation}\label{hada1}
\lambda_{1}=e^{i\frac{\pi}{4}},\qquad \lambda_{2}=e^{i\frac{3\pi}{4}},\qquad \lambda_{3}=e^{i\frac{5\pi}{4}},\qquad \lambda_{4}=e^{i\frac{7\pi}{4}}.
\end{equation}
We prepare the following subsets $K_{1}$, $K_{2}$, and $K_{3}$ of $K=[0,2\pi)$.
\[
K_{1}=\left\{\frac{\pi}{4},\frac{3\pi}{4},\frac{5\pi}{4},\frac{7\pi}{4}\right\},\qquad K_{2}=[0, \pi/4)\cup(3\pi/4, 5\pi/4)\cup(7\pi/4, 2\pi),
\]
\[
K_{3}=K\setminus(K_{1}\cup K_{2}).
\]
From Eq. \eqref{trans2}, the transfer matrices of the Hadamard walk are given by
\begin{align*}
T^+_{\lambda}(H)=
\begin{bmatrix}\vspace{0.2cm}
\displaystyle\frac{2\lambda^2-1}{\sqrt{2}\lambda}&\displaystyle\frac{1}{\sqrt{2}\lambda}\\
\displaystyle\frac{1}{\sqrt{2}\lambda}&\displaystyle-\frac{1}{\sqrt{2}\lambda}
\end{bmatrix},\qquad 
T^{-}_{\lambda}(H)=
\begin{bmatrix}\vspace{0.2cm}
\displaystyle\frac{1}{\sqrt{2}\lambda}&\displaystyle\frac{1}{\sqrt{2}\lambda}\\
\displaystyle\frac{1}{\sqrt{2}\lambda}&\displaystyle-\frac{2\lambda^2-1}{\sqrt{2}\lambda}
\end{bmatrix}.
\end{align*}
{\bf{Remark\ 1.}}
We put $\lambda=e^{i\theta}\in S^{1}$ in Eq. \eqref{eig.pro}, where $\theta\in K$. The transfer matrices $T^{+}_{\lambda}(H)$, $T^{-}_{\lambda}(H)$ are a unitary matrix if and only if $\theta=0,\pi$. For any $\Psi(0)\in\mathbb{C}^{2}\setminus\{\zerovec\}$, we define the function $\Psi\in\mbox{Map}(\mathbb{Z},\mathbb{C}^{2})$
\begin{eqnarray*}
\Psi(x)=
\begin{cases}
\displaystyle (T^{+}_{\lambda}(H))^{x} \Psi(0)&(x\geq 1),\\
\Psi(0)&(x=0),\\
\displaystyle (T^{-}_{\lambda}(H))^{|x|} \Psi(0)&(x\leq -1).
\end{cases}
\end{eqnarray*}
Thus we have
\begin{equation*}
\phi(\Psi)\in\mathcal{M}_{unif}(U_{H})\subset\mathcal{M}_{s,bdd}(U_{C}).
\end{equation*}
Thus, there exist stationary measures in $\mathcal{M}_{s,bdd}(U_{H})$ that has a periodicity. Namely,
\[
\mathcal{M}_{s,period}(U_{H})\cap\mathcal{M}_{s,bdd}(U_{H})\ne\emptyset.
\]
More preciously, we discuss the stationary measures with periodicity in Sec. \ref{type2} Theorem \ref{periodthm}.
\vspace{0.3cm}

\noindent{\bf{Remark\ 2.}}
In \cite{kls}, it is mentioned that the following function $\Psi_{0}^{(\sigma,\tau)}\in\mbox{Map}(\mathbb{Z},\mathbb{C}^{2})\setminus\{\zerovec\}$ satisfies the eigenvalue problem, i.e., there exists $\lambda\in S^{1}$ such that $U_{H}\Psi_{0}^{(\sigma,\tau)}=\lambda\Psi_{0}^{(\sigma,\tau)}$. For $\sigma$, $\tau\in\{\pm1\}$, the function $\Psi_{0}^{(\sigma,\tau)}$ is defined by 
\[
\Psi_{0}^{(\sigma,\tau)}(x)=(\tau i\mbox{sgn}(x))^{|x|}\times
\begin{cases}
\varphi_{1}\times\begin{bmatrix}
1\\
-\sigma\tau i
\end{bmatrix}
&(x\geq1)\\
\begin{bmatrix}
\varphi_{1}\\
\varphi_{2}
\end{bmatrix}
&(x=0)\\
\varphi_{2}\times\begin{bmatrix}
\sigma\tau i\\
1
\end{bmatrix}
&(x\leq1)\\
\end{cases}\qquad(\varphi_{1},\varphi_{2}\in\mathbb{C}).
\]
where $\varphi_{1}=\sigma\tau i\varphi_{2}$ and $\mbox{sgn}(x)$ is given by
\[
\mbox{sgn}(x)=
\begin{cases}
1&(x>0)\\
0&(x=0)\\
-1&(x<0)
\end{cases}.
\]
Then we can check the following equations.
\[
U_{H}\Psi_{0}^{(\sigma,\tau)}=\frac{\sigma+\tau i}{\sqrt{2}}\Psi_{0}^{(\sigma,\tau)},\qquad
\phi\left(\Psi_{0}^{(\sigma,\tau)}\right)\in\mathcal{M}_{unif}(U_{H}).
\]
From now on, we consider the relationship between the transfer matrices $T^{\pm}_{\lambda}(H)$ and the function $\Psi_{0}^{(\sigma,\tau)}$. For simplicity, we take $\sigma$$=$$\tau$$=$$1$, $\varphi_{1}=1$ and $\varphi_{2}$$=$$-i$. Then we have
\[
T^{+}_{\lambda}(H)\Psi_{0}^{(1,1)}(x)=e^{i\frac{\pi}{4}}\ \Psi_{0}^{(1,1)}(x),\qquad T^{-}_{\lambda}(H)\Psi_{0}^{(1,1)}(x)=e^{i\frac{\pi}{4}}\ \Psi_{0}^{(1,1)}(x).
\]
Hence, we see that the function $\Psi_{0}^{(1,1)}\in\mbox{Map}(\mathbb{Z},\mathbb{C}^{2})$ is an eigenfunction of the transfer matrices $T^{+}_{\lambda}(H)$ and $T^{-}_{\lambda}(H)$. Therefore, we conclude that this function $\Psi_{0}^{(1,1)}$ is one of the example that even if there exist $\Psi\in\mbox{Map}(\mathbb{Z},\mathbb{C}^{2})$ such that $\phi(\Psi)\in\mathcal{M}_{unif}(U_{H})$ and $T^{\pm}_{\lambda}(H)\Psi=e^{i\frac{\pi}{4}}\Psi$, the transfer matrices $T^{+}_{\lambda}(H)$ and $T^{-}_{\lambda}(H)$ induced by the eigenvalue $\lambda=e^{i\frac{\pi}{4}}$ are not unitary matrix. Furthermore, we can also obtain the above statement for $\lambda=e^{i\frac{3\pi}{4}}$, $\lambda=e^{i\frac{5\pi}{4}}$, and $\lambda=e^{i\frac{7\pi}{4}}$ by the same discussion.  
\subsubsection{Result of Type $1$}
In Sect. 3, we introduced transfer matrices $T^{+}_{\lambda}(C)$, $T^{-}_{\lambda}(C)$ to obtain stationary measures of quantum walks for Type $1$, $2$, and $3$. In this subsection, by using Theorem \ref{kawaithm}, we present that the stationary measures induced by a solution of the eigenvalue problem of Type $1$, $U_{H}\Psi= \lambda\Psi$, are the stationary measure with quadratic polynomial type.
\begin{proposition}\label{prop1}
Let $\lambda\in S^{1}$ be an eigenvalue in Eq. \eqref{eig.pro} and we put $\lambda=e^{i\theta}$ $(\theta\in K)$. Then we have the following two statements.
\begin{itemize}
\item [$(1)$] The points $\lambda\in S^{1}$ that the characteristic polynomial defined by Eq. \eqref{cheq} has the double roots $\Lambda_{+}=\Lambda_{-}$ are given by
\[
\lambda_{1}=e^{i\frac{\pi}{4}},\qquad \lambda_{2}=e^{i\frac{3\pi}{4}},\qquad \lambda_{3}=e^{i\frac{5\pi}{4}},\qquad \lambda_{4}=e^{i\frac{7\pi}{4}}.
\]
\item [$(2)$]
\begin{itemize}
\item [$(a)$] Suppose that $|\varphi_{1}|^{2}+|\varphi_{2}|^{2}\ne2\Im(\varphi_{1}\overline{\varphi_{2}})$ for $\theta=\pi/4,\ 5\pi/4$. The stationary measures $\phi(\Psi)$ induced by the function $\Psi\in\mbox{Map}(\mathbb{Z},\mathbb{C}^{2})$ in Theorem \ref{kawaithm} $(i)$ have the measures with quadratic polynomial type. That is to say,
\[
\phi(\Psi)\in\mathcal{M}_{s,qp}(U_{H}).
\]
On the other hand, assume that $|\varphi_{1}|^{2}+|\varphi_{2}|^{2}=2\Im(\varphi_{1}\overline{\varphi_{2}})$ for $\theta=\pi/4,\ 5\pi/4$. The stationary measures $\phi(\Psi)$ induced by the function $\Psi\in\mbox{Map}(\mathbb{Z},\mathbb{C}^{2})$ in Theorem \ref{kawaithm} $(i)$ have the measures with period $1$. That is to say,
\[
\phi(\Psi)\in\mathcal{M}_{unif}(U_{H}).
\]

\item [$(b)$] Suppose that $|\varphi_{1}|^{2}+|\varphi_{2}|^{2}\ne-2\Im(\varphi_{1}\overline{\varphi_{2}})$ for $\theta=3\pi/4,\ 7\pi/4$. The stationary measures $\phi(\Psi)$ induced by the function $\Psi\in\mbox{Map}(\mathbb{Z},\mathbb{C}^{2})$ in Theorem \ref{kawaithm} $(i)$ have the measures with quadratic polynomial type. That is to say,
\[
\phi(\Psi)\in\mathcal{M}_{s,qp}(U_{H}).
\]
On the other hand, assume that $|\varphi_{1}|^{2}+|\varphi_{2}|^{2}=-2\Im(\varphi_{1}\overline{\varphi_{2}})$ for $\theta=3\pi/4,\ 7\pi/4$. The stationary measures $\phi(\Psi)$ induced by the function $\Psi\in\mbox{Map}(\mathbb{Z},\mathbb{C}^{2})$ in Theorem \ref{kawaithm} $(i)$ have the measures with period $1$. That is to say,
\[
\phi(\Psi)\in\mathcal{M}_{unif}(U_{H}).
\]
\end{itemize}
\end{itemize}
\end{proposition}
{\bf{Proof.}}
From Eq. \eqref{hada1}, the statement $(1)$ immediately holds. So we show the statement $(2)$. Let $x\in\mathbb{Z}$ be $x\geq1$.  Since $0,\pi\notin K_{1}$, we remark that $\lambda^{2}+\Delta\ne0$. Then we have
\begin{equation}\label{qptf1}
\begin{split}
&2|\Psi^{L}(x)|^{2}\\
&=\left(|\varphi_{1}|^{2}+|\varphi_{2}|^{2}+\varphi_{1}\overline{\varphi_{2}}\lambda^{2}+\overline{\varphi_{1}}\varphi_{2}\overline{\lambda^{2}}\right)x^{2}\\
&+\left(2|\varphi_{1}|^{2}-(\varphi_{1}\overline{\varphi_{2}}+\overline{\varphi_{1}}\varphi_{2})
+\varphi_{1}\overline{\varphi_{2}}\lambda^{2}+\overline{\varphi_{1}}\varphi_{2}\overline{\lambda^{2}}-|\varphi_{1}|^{2}\lambda^{2
}-|\varphi_{1}|^{2}\overline{\lambda^{2}}\right)x\\
&+2|\varphi_{1}|^{2}-|\varphi_{1}|^{2}\lambda^{2
}-|\varphi_{1}|^{2}\overline{\lambda^{2}}.
\end{split}
\end{equation}
\begin{equation}\label{qptf2}
\begin{split}
&2|\Psi^{R}(x)|^{2}\\
&=\left(|\varphi_{1}|^{2}+|\varphi_{2}|^{2}-\varphi_{1}\overline{\varphi_{2}}\overline{\lambda^{2}}-\overline{\varphi_{1}}\varphi_{2}\lambda^{2}\right)x^{2}\\
&+\left(-2|\varphi_{2}|^{2}-(\varphi_{1}\overline{\varphi_{2}}+\overline{\varphi_{1}}\varphi_{2})
+\varphi_{1}\overline{\varphi_{2}}\overline{\lambda^{2}}+\overline{\varphi_{1}}\varphi_{2}\lambda^{2}+|\varphi_{2}|^{2}\overline{\lambda^{2
}}+|\varphi_{2}|^{2}\lambda^{2}\right)x\\
&+2|\varphi_{2}|^{2}-|\varphi_{2}|^{2}\lambda^{2
}-|\varphi_{2}|^{2}\overline{\lambda^{2}}.
\end{split}
\end{equation}
Thus, we obtain the following stationary measure by using Eqs. \eqref{qptf1} and \eqref{qptf2}.
\begin{equation*}
\begin{split}
\mu(x)&=|\Psi^{L}(x)|^2+|\Psi^{R}(x)|^2\\
&=\left(|\varphi_{1}|^{2}+|\varphi_{2}|^{2}-2\sin2\theta\Im(\varphi_{1}\overline{\varphi_{2}})\right)x^{2}\\
&\hspace{1cm}+\left(|\varphi_{1}|^{2}-|\varphi_{2}|^{2}-2\Re(\varphi_{1}\overline{\varphi_{2}})\right)x
+|\varphi_{1}|^{2}+|\varphi_{2}|^{2}.
\end{split}
\end{equation*}
Suppose that the condition $|\varphi_{1}|^{2}+|\varphi_{2}|^{2}\ne2\Im(\varphi_{1}\overline{\varphi_{2}})$ for $\theta=\pi/4,\ 5\pi/4$ and assume that $|\varphi_{1}|^{2}+|\varphi_{2}|^{2}\ne-2\Im(\varphi_{1}\overline{\varphi_{2}})$ for $\theta=3\pi/4,\ 7\pi/4$ $(\varphi_{1},\varphi_{2}\in\mathbb{C})$. Then it holds
\[
\mu\in\mathcal{M}_{s,qp}(U_{H}).
\]
Moreover, note that
\[
|\varphi_{1}|^{2}+|\varphi_{2}|^{2}=2\Im(\varphi_{1}\overline{\varphi_{2}})\Longrightarrow
|\varphi_{1}|=|\varphi_{2}|,\ \Re(\varphi_{1}\overline{\varphi_{2}})=0
\]
and
\[
|\varphi_{1}|^{2}+|\varphi_{2}|^{2}=-2\Im(\varphi_{1}\overline{\varphi_{2}})\Longrightarrow
|\varphi_{1}|=|\varphi_{2}|,\ \Re(\varphi_{1}\overline{\varphi_{2}})=0.
\]
Suppose that the condition $|\varphi_{1}|^{2}+|\varphi_{2}|^{2}=2\Im(\varphi_{1}\overline{\varphi_{2}})$ for $\theta=\pi/4,\ 5\pi/4$ and assume that $|\varphi_{1}|^{2}+|\varphi_{2}|^{2}=-2\Im(\varphi_{1}\overline{\varphi_{2}})$ for $\theta=3\pi/4,\ 7\pi/4$. Then we obtain
\[
\mu\in\mathcal{M}_{unif}(U_{H}).
\]
In case of $x\in\mathbb{Z}$ with $x\leq-1$, we get the same results by the same argument. Hence the statement $(2)$ holds. 
\hspace{\fill} $\square$
\vspace{0.3cm}

From the above argument, we have obtained an explicit necessary and sufficient condition to have the uniform measures for $\theta\in K_{1}$.
\begin{corollary}
Let $\mu$ be a measure given by
\[
\mu(x)=\begin{cases}
\displaystyle \|(T^{+}_{\lambda}(H))^{x} \varphi\|^{2}_{\mathbb{C}^{2}}&(x\geq 1)\\
\|\varphi\|^{2}_{\mathbb{C}^{2}}&(x=0)\\
\displaystyle \|(T^{-}_{\lambda}(H))^{|x|} \varphi\|^{2}_{\mathbb{C}^{2}}&(x\leq -1)
\end{cases}.
\]
For $\theta\in K_{1}$, we have the following results.
\begin{itemize}
\item [$(1)$] For $\theta=\displaystyle\frac{\pi}{4},\ \frac{5\pi}{4}$, we obtain
\[
\mu\in\mathcal{M}_{unif}(U_{H})\Longleftrightarrow\varphi\in S^{(1)}_{unif},
\]
where $S^{(1)}_{unif}$ is given by 
\[
S^{(1)}_{unif}=\left\{\varphi=\begin{bmatrix}
\varphi_{1}\\
\varphi_{2}
\end{bmatrix}\in\mathbb{C}^{2}:|\varphi_{1}|=|\varphi_{2}|, \arg(\varphi_{1})-\arg(\varphi_{2})=\frac{\pi}{2}+2n\pi\ (n\in\mathbb{Z})\right\}.
\]
\item [$(2)$] For $\theta=\displaystyle\frac{3\pi}{4},\ \frac{7\pi}{4}$, we obtain
\[
\mu\in\mathcal{M}_{unif}(U_{H})\Longleftrightarrow\varphi\in S^{(2)}_{unif},
\]
where $S^{(2)}_{unif}$ is given by 
\[
S^{(2)}_{unif}=\left\{\varphi=\begin{bmatrix}
\varphi_{1}\\
\varphi_{2}
\end{bmatrix}\in\mathbb{C}^{2}:|\varphi_{1}|=|\varphi_{2}|, \arg(\varphi_{1})-\arg(\varphi_{2})=\frac{3\pi}{2}+2n\pi\ (n\in\mathbb{Z})\right\}.
\]
\end{itemize}
\end{corollary}
\subsubsection{Result of Type $2$}\label{type2}
In the previous subsection, we determined the points $\lambda\in S^{1}$ that the characteristic polynomial defined by Eq. \eqref{cheq} has the double roots. After that we showed that the stationary measures of Type $1$ are measures with quadratic polynomial type. This subsection deals with the stationary measures of Type $2$ . We prepare the following lemma to prove Proposition \ref{prop2} and Proposition \ref{prop3}.
\begin{lemma}\label{fg1}
Let $f(\lambda)$ and $g(\lambda)$ be the following functions on $S^{1}$. 
\[
f(\lambda)=\lambda^{2}-1,\qquad g(\lambda)=\sqrt{\lambda^{4}+1},
\]
and we define $z(\lambda)$ as
\[
z(\lambda)=\overline{f(\lambda)}g(\lambda).
\]
\begin{itemize}
\item [$(1)$] For $\theta\in K_{2}$, we have the followings.
\vspace{0.1cm}
\begin{itemize}
\item [$(i)$] $\Re(z(\lambda))=0.$
\vspace{0.3cm}
\item [$(ii)$] $\begin{cases}
\Im(z(\lambda))=-2\sin\theta\sqrt{2\cos2\theta}  & \left(0\leq\theta<\frac{\pi}{4},\ \frac{3\pi}{4}<\theta\leq\pi\right),\\
\Im(z(\lambda))=2\sin\theta\sqrt{2\cos2\theta}  & \left(\pi\leq\theta<\frac{5\pi}{4},\ \frac{7\pi}{4}<\theta<2\pi\right).
\end{cases}
$
\end{itemize}
\vspace{0.3cm}
\item [$(2)$] For $\theta\in K_{3}$, we have the followings.
\vspace{0.1cm}
\begin{itemize}
\item [$(i)$] 
$
\begin{cases}
\Re(z(\lambda))=2\sin\theta\sqrt{-2\cos2\theta}&\left(\frac{\pi}{4}<\theta<\frac{\pi}{2},\ \frac{3\pi}{2}\leq\theta<\frac{7\pi}{4}\right),\\
\Re(z(\lambda))=-2\sin\theta\sqrt{-2\cos2\theta}&\left(\frac{\pi}{2}\leq\theta<\frac{3\pi}{4},\ \frac{5\pi}{4}<\theta<\frac{3\pi}{2}\right).
\end{cases}
$\vspace{0.3cm}
\item [$(ii)$] $\Im(z(\lambda))=0.$
\end{itemize}
\end{itemize}
\end{lemma}
{\bf{Proof.}}
We define the function $z(\lambda)$ on $S^{1}$ as
\[
z(\lambda)=\overline{f(\lambda)}g(\lambda).
\]
Let $t=\cos2\theta$. It holds that
\[
t=0\Longleftrightarrow \theta\in K_{1},\qquad0<t\leq1\Longleftrightarrow \theta\in K_{2},\qquad -1\leq t<0\Longleftrightarrow \theta\in K_{3}.
\]
We set $g(\lambda)=re^{i\eta}\ (r\ne0)$. Then we get the following relation.
\begin{equation}\label{re11}
r^{2}\cos2\eta=2\cos^{2}2\theta,\qquad r^{2}\sin2\eta=2\sin2\theta\cos2\theta.
\end{equation}
From the above Eq. \eqref{re11}, it holds
\[
\eta=\theta+\frac{\pi}{2}n\qquad(n\in\mathbb{Z}).
\]
Hence, we have
\[
\begin{cases}
r=\sqrt{2\cos2\theta}&(\theta\in K_{2}),\\
r=\sqrt{-2\cos2\theta}&(\theta\in K_{3}).
\end{cases}
\]
Note that
\begin{equation*}
\begin{split}
z(\lambda)&=r(e^{-2i\theta}-1)e^{i\eta}\\
&=r\{\cos(2\theta-\eta)-\cos\eta\}+ir\{\sin(\eta-2\theta)-\sin\eta\}.
\end{split}
\end{equation*}
Then we have
\begin{equation*}
\begin{split}
\Re(z(\lambda))&=r\{\cos(2\theta-\eta)-\cos\eta\}\\
&=2r\sin\theta\sin\frac{\pi}{2}n\\
&=
\begin{cases}
 0 & \hspace{-4cm}(\theta \in K_{2}),\\
          \\
\begin{cases}
2\sin\theta\sqrt{-2\cos2\theta}  & \left(\frac{\pi}{4}<\theta<\frac{\pi}{2},\ \frac{3\pi}{2}\leq\theta<\frac{7\pi}{4}\right),\\
-2\sin\theta\sqrt{-2\cos2\theta}  & \left(\frac{\pi}{2}\leq\theta<\frac{3\pi}{4},\ \frac{5\pi}{4}<\theta<\frac{3\pi}{2}\right).
 \end{cases}
      \end{cases}
\end{split}
\end{equation*}
On the other hand, we obtain
\begin{equation*}
\begin{split}
\Im(z(\lambda))&=r\{\sin(\eta-2\theta)-\sin\eta\}\\
&=-2r\sin\theta\cos\frac{\pi}{2}n\\
&=
\begin{cases}
 0 & \hspace{-4cm}(\theta \in K_{3}),\\
          \\
\begin{cases}
-2\sin\theta\sqrt{2\cos2\theta}  & \left(0\leq\theta<\frac{\pi}{4},\ \frac{3\pi}{4}<\theta\leq\pi\right),\\
2\sin\theta\sqrt{2\cos2\theta}  & \left(\pi\leq\theta<\frac{5\pi}{4},\ \frac{7\pi}{4}<\theta<2\pi\right),
      \end{cases}
      \end{cases}
\end{split}
\end{equation*}
which shows the statements $(1)$ and $(2)$.
\hspace{\fill} $\square$
\vspace{0.3cm}
\begin{proposition}\label{prop2}
Let $\lambda\in S^{1}$ be an eigenvalue in Eq. \eqref{eig.pro} and we put $\lambda=e^{i\theta}$ $(\theta\in K)$. Suppose that $\theta\in K_{2}$. Then we have the following two statements.
\begin{itemize}
\item [$(1)$] $|\Lambda_{+}|=|\Lambda_{-}|$ $(\Lambda_{+}\ne\Lambda_{-})$.
\item [$(2)$] Suppose that $|\varphi_{1}|$, $|\varphi_{2}|<\infty$. The stationary measures $\phi(\Psi)$ induced by the function $\Psi\in\mbox{Map}(\mathbb{Z},\mathbb{C}^{2})$ in Theorem \ref{kawaithm} $(ii)$ have the measures with bounded type. That is to say,
\[
\phi(\Psi)\in\mathcal{M}_{s,bdd}(U_{H}).
\]
\end{itemize}
\end{proposition}
{\bf{Proof.}}
At first, we show the statement $(1)$. From Theorem \ref{kawaithm}, it holds that
\[
\Lambda_{\pm}=\frac{f(\lambda)\pm g(\lambda)}{\sqrt{2}\lambda}.
\]
From Lemma \ref{fg1}, we can compute $|\Lambda_{+}|^{2}$ and $|\Lambda_{-}|^{2}$ as
\begin{equation}\label{lam+}
\begin{split}
|\Lambda_{+}|^{2}&=\frac{1}{2}\left\{(\lambda^{2}-1)+\sqrt{\lambda^{4}+1}\right\}\left\{(\overline{\lambda^{2}}-1)+\overline{\sqrt{\lambda^{4}+1}}\right\}\\
&=\frac{1}{2}\left\{|f(\lambda)|^{2}+|g(\lambda)|^{2}+2\Re(z(\lambda))\right\}\\
&=\begin{cases}
 1 & \hspace{-4cm}(\theta \in K_{2}),\\
          \\
\begin{cases}
 1-2\cos2\theta+2\sin\theta\sqrt{-2\cos2\theta}  & \left(\frac{\pi}{4}<\theta<\frac{\pi}{2},\ \frac{3\pi}{2}\leq\theta<\frac{7\pi}{4}\right),\\
  1-2\cos2\theta-2\sin\theta\sqrt{-2\cos2\theta} & \left(\frac{\pi}{2}\leq\theta<\frac{3\pi}{4},\ \frac{5\pi}{4}<\theta<\frac{3\pi}{2}\right)
 \end{cases}
      \end{cases}
\end{split}
\end{equation}
and 
\begin{equation}\label{lam-}
\begin{split}
|\Lambda_{-}|^{2}&=\frac{1}{2}\left\{(\lambda^{2}-1)-\sqrt{\lambda^{4}+1}\right\}\left\{(\overline{\lambda^{2}}-1)-\overline{\sqrt{\lambda^{4}+1}}\right\}\\
&=\frac{1}{2}\left\{|f(\lambda)|^{2}+|g(\lambda)|^{2}-2\Re(z(\lambda))\right\}\\
&=\begin{cases}
 1 & \hspace{-4cm}(\theta \in K_{2}),\\
          \\
\begin{cases}
1-2\cos2\theta-2\sin\theta\sqrt{-2\cos2\theta}  & \left(\frac{\pi}{4}<\theta<\frac{\pi}{2},\ \frac{3\pi}{2}\leq\theta<\frac{7\pi}{4}\right),\\
  1-2\cos2\theta+2\sin\theta\sqrt{-2\cos2\theta}   & \left(\frac{\pi}{2}\leq\theta<\frac{3\pi}{4},\ \frac{5\pi}{4}<\theta<\frac{3\pi}{2}\right).
 \end{cases}
      \end{cases}
\end{split}
\end{equation}
By using Eqs. \eqref{lam+} and \eqref{lam-}, we obtain
\[
|\Lambda_{+}|=|\Lambda_{-}|\qquad(\theta\in K_{2}).
\]
Since $g(\lambda)\ne0$ for $\theta\in K_{2}$, we remark that $\Lambda_{+}\ne\Lambda_{-}$. Thus, the statement $(1)$ holds. Next, we show the statement $(2)$. Let $x\in\mathbb{Z}$ with $x\geq1$. We note the following.
\[
|\Lambda_{+}|=|\Lambda_{-}|=1\ (\theta\in K_{2}),\qquad\left|\frac{1}{\Lambda_{+}-\Lambda_{-}}\right|^{2}
=\begin{cases}
\displaystyle\frac{1}{4\cos2\theta} & (\theta \in K_{2}),\\
          \\
\displaystyle\frac{1}{-4\cos2\theta}  & (\theta \in K_{3}).
      \end{cases}
\]
Furthermore, we obtain the following formula about $\Lambda_{+}\cdot\overline{\Lambda_{-}}$.
\begin{equation*}
\begin{split}
\Lambda_{+}\cdot\overline{\Lambda_{-}}&=\frac{1}{2}\left\{(\lambda^{2}-1)+\sqrt{\lambda^{4}+1}\right\}\left\{(\overline{\lambda^{2}}-1)-\overline{\sqrt{\lambda^{4}+1}}\right\}\\
&=\frac{1}{2}\left\{|f(\lambda)|^{2}-|g(\lambda)|^{2}+2i\Im(z(\lambda))\right\}\\
&=\begin{cases}
 1  & \hspace{-4cm}(\theta \in K_{3}),
 \vspace{0.4cm}
          \\
          \begin{cases}
 1-2\cos2\theta- i2\sqrt{2\cos2\theta}\sin\theta  & \left(0\leq\theta<\frac{\pi}{4},\ \frac{3\pi}{4}<\theta\leq\pi\right),\\
 1-2\cos2\theta+i2\sqrt{2\cos2\theta}\sin\theta  & \left(\pi\leq\theta<\frac{5\pi}{4},\ \frac{7\pi}{4}<\theta<2\pi\right).\\
 \end{cases}
      \end{cases}
\end{split}
\end{equation*}
In case of $\theta\in K_{2}$, there exists $\xi_{j}\in(0,2\pi)$ such that
\[
\Lambda_{+}\cdot\overline{\Lambda_{-}}=
\begin{cases}
e^{i\xi_{1}}&\left(0\leq\theta<\frac{\pi}{4},\ \frac{3\pi}{4}<\theta\leq\pi\right),\\
e^{i\xi_{2}}&\left(\pi\leq\theta<\frac{5\pi}{4},\ \frac{7\pi}{4}<\theta<2\pi\right).
\end{cases}
\]
Here, $\xi_{1}$ and $\xi_{2}$ are defined by
\begin{equation}\label{xi2}
\cos\xi_{1}=1-2\cos2\theta,\qquad\sin\xi_{1}=-2\sqrt{2\cos2\theta}\sin\theta ,\qquad 
e^{i\xi_{2}}=e^{-i\xi_{1}}.
\end{equation}
From now on, we consider $\theta\in K_{2}$. By using Theorem \ref{kawaithm}, it holds that
\begin{equation*}
\begin{split}
|\Psi^{L}(x)|^{2}&=\left|\frac{1}{\Lambda_{+}-\Lambda_{-}}\right|^{2}\Biggl\{|h_{1}|^{2}+|h_{2}|^{2}-2\Re\left((\Lambda_{+}\cdot\overline{\Lambda_{-}})^{x}h_{1}\overline{h_{2}}\right)\Biggr\},\\
\end{split}
\end{equation*}
where $h_{1}$ and $h_{2}$ are given by
\begin{equation}\label{h1}
h_{1}=\Psi^{L}(1)-\Lambda_{-}\varphi_{1},\qquad h_{2}=\Psi^{L}(1)-\Lambda_{+}\varphi_{1}.
\end{equation}
Furthermore, $|\Psi^{R}(x)|^{2}$ is computed  as
\begin{equation*}
\begin{split}
|\Psi^{R}(x)|^{2}&=\left|\frac{1}{\Lambda_{+}-\Lambda_{-}}\right|^{2}\Biggl\{|h_{3}|^{2}+|h_{4}|^{2}-2\Re\left((\Lambda_{+}\cdot\overline{\Lambda_{-}})^{x}h_{3}\overline{h_{4}}\right)\Biggr\},\\
\end{split}
\end{equation*}
where $h_{3}$ and $h_{4}$ are given by
\begin{equation}\label{h3}
h_{3}=\Psi^{R}(1)-\Lambda_{-}\varphi_{2},\qquad h_{4}=\Psi^{R}(1)-\Lambda_{+}\varphi_{2}.
\end{equation}
Therefore, we have
\begin{equation*}
\begin{split}
\mu(x)&=|\Psi^{L}(x)|^{2}+|\Psi^{R}(x)|^{2}\\
&=\left|\frac{1}{\Lambda_{+}-\Lambda_{-}}\right|^{2}\Biggl\{\sum_{i=1}^{4}|h_{i}|^{2}-2\Re\left((\Lambda_{+}\cdot\overline{\Lambda_{-}})^{x}(h_{1}\overline{h_{2}}+h_{3}\overline{h_{4}})\right)\Biggr\}\\
&=\left|\frac{1}{\Lambda_{+}-\Lambda_{-}}\right|^{2}\Biggl\{W_{1}(\varphi_{1},\varphi_{2},\theta)-2\Re\left((\Lambda_{+}\cdot\overline{\Lambda_{-}})^{x}W_{2}(\varphi_{1},\varphi_{2},\theta)\right)\Biggr\}.
\end{split}
\end{equation*}
Here, $W_{1}(\varphi_{1},\varphi_{2},\theta)$ and $W_{2}(\varphi_{1},\varphi_{2},\theta)$ are defined by
\[
W_{1}(\varphi_{1},\varphi_{2},\theta)=\sum_{i=1}^{4}|h_{i}|^{2},\qquad 
W_{2}(\varphi_{1},\varphi_{2},\theta)=h_{1}\overline{h_{2}}+h_{3}\overline{h_{4}}.
\]
We put $k_{1}$ and $k_{2}$ as
\begin{equation*}
k_{1}=\Psi^{L}(-1)-\Gamma_{-}\varphi_{1},\qquad k_{2}=\Psi^{L}(-1)-\Gamma_{+}\varphi_{1}
\end{equation*}
and $k_{3}$ and $k_{4}$ are given by
\begin{equation*}
k_{3}=\Psi^{R}(-1)-\Gamma_{-}\varphi_{2},\qquad k_{4}=\Psi^{R}(-1)-\Gamma_{+}\varphi_{2}.
\end{equation*}
For $x\leq-1$, we have
\begin{equation*}
\begin{split}
\mu(x)&=|\Psi^{L}(x)|^{2}+|\Psi^{R}(x)|^{2}\\
&=\left|\frac{1}{\Lambda_{+}-\Lambda_{-}}\right|^{2}\Biggl\{\sum_{i=1}^{4}|k_{i}|^{2}-2\Re\left((\Lambda_{+}\cdot\overline{\Lambda_{-}})^{x}(k_{1}\overline{k_{2}}+k_{3}\overline{k_{4}})\right)\Biggr\}\\
&=\left|\frac{1}{\Lambda_{+}-\Lambda_{-}}\right|^{2}\Biggl\{W_{3}(\varphi_{1},\varphi_{2},\theta)-2\Re\left((\Lambda_{+}\cdot\overline{\Lambda_{-}})^{x}W_{4}(\varphi_{1},\varphi_{2},\theta)\right)\Biggr\}.
\end{split}
\end{equation*}
Here, $W_{3}(\varphi_{1},\varphi_{2},\theta)$ and $W_{4}(\varphi_{1},\varphi_{2},\theta)$ are defined by
\[
W_{3}(\varphi_{1},\varphi_{2},\theta)=\sum_{i=1}^{4}|k_{i}|^{2},\qquad 
W_{4}(\varphi_{1},\varphi_{2},\theta)=k_{1}\overline{k_{2}}+k_{3}\overline{k_{4}}.
\]
We set $W_{j}(\varphi_{1},\varphi_{2},\theta)=r_{j}e^{i\eta_{j}}\in\mathbb{C}$, where $r_{j}\geq0$ and $j=2,4$. By using the assumption $|\varphi_{1}|$, $|\varphi_{2}|<\infty$, we obtain
\[
 W_{i}(\varphi_{1},\varphi_{2},\theta)<\infty\ (i=1,3), \qquad 0\leq r_{j}<\infty\ (j=2,4).
\]
Therefore, we have $\mu(x)<\infty$. Namely, 
\[
\mu\in\mathcal{M}_{s,bdd}(U_{H}).
\]
\hspace{\fill} $\square$
\vspace{0.3cm}

\noindent{\bf{Remark\ 3.}}
For $\theta=0$ and $\pi$, it holds 
\[
W_{2}(\varphi_{1},\varphi_{2},\theta)=0,\qquad W_{4}(\varphi_{1},\varphi_{2},\theta)=0\qquad \left(\varphi=\begin{bmatrix}\varphi_{1}\\ \varphi_{2}\\ \end{bmatrix}\in\mathbb{C}^{2}\right).
\]
\noindent{\bf{Remark\ 4.}}
If $W_{2}(\varphi_{1},\varphi_{2},\theta)=0$, we obtain
\begin{equation*}
\begin{split}
W_{1}(\varphi_{1},\varphi_{2},\theta)&=(h_{1}-h_{2})(\overline{h_{1}}-\overline{h_{2}})+(h_{3}-h_{4})(\overline{h_{3}}-\overline{h_{4}})\\
&=4\cos2\theta(|\varphi_{1}|^{2}+|\varphi_{2}|^{2})\ne0.
\end{split}
\end{equation*}
If $W_{4}(\varphi_{1},\varphi_{2},\theta)=0$, we get
\begin{equation*}
\begin{split}
W_{3}(\varphi_{1},\varphi_{2},\theta)&=(k_{1}-k_{2})(\overline{k_{1}}-\overline{k_{2}})+(k_{3}-k_{4})(\overline{k_{3}}-\overline{k_{4}})\\
&=4\cos2\theta(|\varphi_{1}|^{2}+|\varphi_{2}|^{2})\ne0.
\end{split}
\end{equation*}
\begin{theorem}\label{periodthm}
Suppose that $\theta\in K_{2}$. Then it holds the following two statements.
\begin{itemize}
\item [$(1)$] Assume that $W_{2}(\varphi_{1},\varphi_{2},\theta)=0$ and $W_{4}(\varphi_{1},\varphi_{2},\theta)=0$. Then we have
\[
\mu\in\mathcal{M}_{s,period}^{(1)}(U_{H}).
\]

\item [$(2)$] Assume that $W_{2}(\varphi_{1},\varphi_{2},\theta)\ne0$ and $W_{4}(\varphi_{1},\varphi_{2},\theta)\ne0$. The necessary and sufficient condition for $\xi_{j}\ (j=1,2)$ given in Eq. \eqref{xi2} to have the stationary measure $\mu$ with a periodicity is 
\[
\begin{cases}
\vspace{0.4cm}
\displaystyle\frac{2n\pi}{\xi_{1}}\in\mathbb{N}& \left(0\leq\theta<\displaystyle\frac{\pi}{4},\ \frac{3\pi}{4}<\theta\leq\pi\right),\\
\displaystyle\frac{2n\pi}{\xi_{2}}\in\mathbb{N} & \left(\displaystyle\pi\leq\theta<\displaystyle\frac{5\pi}{4},\ \displaystyle\frac{7\pi}{4}<\theta<\displaystyle2\pi\right).
\end{cases}
\]
Here $n\in\mathbb{N}$.
\end{itemize}
\end{theorem}
{\bf{Proof.}}
For $\mu\in\displaystyle\bigcup_{\lambda\in \widetilde{K_{2}}}\mathcal{M}_{s}^{(\lambda)}(U_{H})$, it holds
\[
\mu(x)=
\begin{cases}
\vspace{0.3cm}
\left|\frac{1}{\Lambda_{+}-\Lambda_{-}}\right|^{2}\Biggl(W_{1}(\varphi_{1},\varphi_{2},\theta)-2\Re\left((\Lambda_{+}\cdot\overline{\Lambda_{-}})^{x}W_{2}(\varphi_{1},\varphi_{2},\theta)\right)\Biggr)&x\geq1,\\
\vspace{0.3cm}
|\varphi_{1}|^{2}+|\varphi_{2}|^{2}&x=0,\\
\left|\frac{1}{\Lambda_{+}-\Lambda_{-}}\right|^{2}\Biggl(W_{3}(\varphi_{1},\varphi_{2},\theta)-2\Re\left((\Lambda_{+}\cdot\overline{\Lambda_{-}})^{x}W_{4}(\varphi_{1},\varphi_{2},\theta)\right)\Biggr)&x\leq-1.

\end{cases}
\]
From the above equation and Remark $4$, the statement $(1)$ holds.  Next, we show the statement $(2)$. Suppose that $W_{2}(\varphi_{1},\varphi_{2},\theta)\ne0$ and $W_{4}(\varphi_{1},\varphi_{2},\theta)\ne0$. We put $W_{2}(\varphi_{1},\varphi_{2},\theta)=re^{i\eta}\in\mathbb{C}\setminus\{0\}$, where $r$ is a positive real number.  For $m\in\mathbb{N}$, we obtain
\[
\begin{split}
&\mu(x+m)=\mu(x)\\
&\Longleftrightarrow \Re\left(\left(\Lambda_{+}\cdot\overline{\Lambda_{-}}\ \right)^{x+m}W_{2}(\varphi_{1},\varphi_{2},\theta)\right)=\Re\left(\left(\Lambda_{+}\cdot\overline{\Lambda_{-}}\ \right)^{x}W_{2}(\varphi_{1},\varphi_{2},\theta)\right)\\
&\Longleftrightarrow 
\Re\left(e^{i\xi_{j}(x+m)}\cdot e^{i\eta}\right)=\Re\left(e^{i\xi_{j} x}\cdot e^{i\eta}\right)\\
&\Longleftrightarrow \cos(x\xi_{j}+\eta+m\xi_{j})=\cos(x\xi_{j}+\eta)\\
&\Longleftrightarrow x\xi_{j}+\eta+m\xi_{j}=x\xi_{j}+\eta+2n\pi\qquad(n\in\mathbb{Z})\\
&\Longleftrightarrow m=\frac{2n\pi}{\xi_{j}}\qquad(n\in\mathbb{Z}).
\end{split}
\] 
In case of $x\in\mathbb{Z}$ with $x \leq-1$, we get the same results by the same argument. Hence, this completes the proof of Theorem \ref{periodthm}.
\hspace{\fill} $\square$
\vspace{0.3cm}

\noindent Let $m_{min}$ be the minimum value satisfied with $\mu(x+m)=\mu(x)\ (x\in\mathbb{Z})$. We call this natural number $m_{min}$ {\it periodicity} to the stationary measure $\mu$.
\vspace{0.3cm}

\noindent{\bf{Example\ 1.}} We consider the case of $\theta=0$ ($\theta=\pi$). Note that $W_{2}(\varphi_{1},\varphi_{2},\theta)=0\ (\theta=0,\pi)$. Then it holds $m_{min}=1$. The stationary measure induced by $U_{H}\Psi=\Psi$ $(U_{H}\Psi=\pi\Psi)$ is satisfied with 
\[
\phi(\Psi)\in\mathcal{M}_{s,period}^{(1)}(U_{H})
\]
\noindent{\bf{Example\ 2.}} We consider the case of $\theta=\pi/6$. Then we obtain
\[
\Lambda_{+}=e^{i\frac{\pi}{4}},\qquad\Lambda_{-}=e^{i\frac{3\pi}{4}},\qquad \Lambda_{+}\cdot\overline{\Lambda_{-}}=e^{i\frac{3\pi}{2}}.
\]
Thus $\xi$ is $3\pi/2$. From Theorem \ref{periodthm}, we have 
\[
m_{min}=\min_{m}\left\{m\in\mathbb{N}\ :\ m=2n\pi\times\frac{2}{3\pi}\ (n\in\mathbb{N})\right\}=4.
\]
Therefore, the stationary measure induced by $U_{H}\Psi=\pi/6\Psi$ for the Hadamard walk  has a period $4$. That is to say,
\[
\phi(\Psi)\in\mathcal{M}_{s,period}^{(4)}(U_{H}).
\]
\subsubsection{Result of Type $3$}
In the previous subsection, we determined the stationary measures of Type $2$ given by the characteristic polynomial for $\theta\in K_{2}$. Moreover, we see that there exists $\theta\in K_{2}$ such that $\phi(\Psi)$ is a stationary measure with periodicity, where $U_{H}\Psi=e^{i\theta}\Psi$. This subsection deals with the stationary measures of Type $3$.
\begin{proposition}\label{prop3}
Let $\lambda\in S^{1}$ be an eigenvalue in Eq. \eqref{eig.pro} and we put $\lambda=e^{i\theta}$ $(\theta\in K)$. Suppose that $\theta\in K_{3}$. Then we have the following two statements.
\begin{itemize}
\item [$(1)$] For $\theta\in K_{3}$, we have
\[
\begin{cases}
\vspace{0.3cm}
|\Lambda_{+}|>1>|\Lambda_{-}|>0\qquad \left(\displaystyle\frac{\pi}{4}<\theta<\displaystyle\frac{\pi}{2},\ \displaystyle\frac{5\pi}{4}<\theta<\displaystyle\frac{3\pi}{2}\right),\\
|\Lambda_{-}|>1>|\Lambda_{+}|>0\qquad \left(\displaystyle\frac{\pi}{2}\leq\theta<\displaystyle\frac{3\pi}{4},\ \displaystyle\frac{3\pi}{2}\leq\theta<\displaystyle\frac{7\pi}{4}\right).

\end{cases}
\]

\item [$(2)$] The stationary measures $\phi(\Psi)$ induced by the function $\Psi\in\mbox{Map}(\mathbb{Z},\mathbb{C}^{2})$ in Theorem \ref{kawaithm} $(ii)$ have the measures with exponential type. That is to say,
\[
\phi(\Psi)\in\mathcal{M}_{s,exp}(U_{H}).
\]
\end{itemize}
\end{proposition} 
{\bf{Proof.}}
From Eqs. \eqref{lam+} and \eqref{lam-}, we obtain
\[
\begin{cases}
\vspace{0.3cm}
|\Lambda_{+}|>1>|\Lambda_{-}|>0\qquad \left(\displaystyle\frac{\pi}{4}<\theta<\displaystyle\frac{\pi}{2},\ \displaystyle\frac{5\pi}{4}<\theta<\displaystyle\frac{3\pi}{2}\right),\\
|\Lambda_{-}|>1>|\Lambda_{+}|>0\qquad \left(\displaystyle\frac{\pi}{2}\leq\theta<\displaystyle\frac{3\pi}{4},\ \displaystyle\frac{3\pi}{2}\leq\theta<\displaystyle\frac{7\pi}{4}\right).

\end{cases}
\]
It holds the statement $(1)$. Next, we show that thae statement $(2)$. Since the proof of Proposition \ref{prop3} $(2)$ under the conditions $|\Lambda_{-}|>1>|\Lambda_{+}|>0$ is the same as that of Proposition \ref{prop3} $(2)$ under the condition $|\Lambda_{+}|>1>|\Lambda_{-}|>0$, we only give the proof of the latter. From Theorem \ref{kawaithm}, it holds that
\begin{equation*}
\begin{split}
|\Psi^{L}(x)|^{2}&=\left|\frac{1}{\Lambda_{+}-\Lambda_{-}}\right|^{2}\Biggl\{(|\Lambda_{+}|^{2})^{x}|h_{1}|^{2}+(|\Lambda_{-}|^{2})^{x}|h_{2}|^{2}-2\Re\left(h_{1}\overline{h_{2}}\right)\Biggr\},\\
\end{split}
\end{equation*}
where $h_{1}$ and $h_{2}$ are given by Eq. \eqref{h1}. Furthermore, $|\Psi^{R}(x)|^{2}$ is computed  as
\begin{equation*}
\begin{split}
|\Psi^{R}(x)|^{2}&=\left|\frac{1}{\Lambda_{+}-\Lambda_{-}}\right|^{2}\Biggl\{(|\Lambda_{+}|^{2})^{x}|h_{3}|^{2}+(|\Lambda_{-}|^{2})^{x}|h_{4}|^{2}-2\Re\left(h_{3}\overline{h_{4}}\right)\Biggr\},\\
\end{split}
\end{equation*}
where $h_{3}$ and $h_{4}$ are given by Eq. \eqref{h3}.
Therefore, we have
\begin{equation*}
\begin{split}
\mu(x)&=|\Psi^{L}(x)|^{2}+|\Psi^{R}(x)|^{2}\\
&=\left|\frac{1}{\Lambda_{+}-\Lambda_{-}}\right|^{2}\Biggl\{(|\Lambda_{+}|^{2})^{x}(|h_{1}|^{2}+|h_{3}|^{2})+(|\Lambda_{-}|^{2})^{x}(|h_{2}|^{2}+|h_{4}|^{2})\\
&\hspace{7cm}-2\Re\left(h_{1}\overline{h_{2}}+h_{3}\overline{h_{4}}\right)\Biggr\}\\
&=\left|\frac{1}{\Lambda_{+}-\Lambda_{-}}\right|^{2}\Biggl\{(|\Lambda_{+}|^{2})^{x}W_{5}(\varphi_{1},\varphi_{2},\theta)+(|\Lambda_{-}|^{2})^{x}W_{6}(\varphi_{1},\varphi_{2},\theta)\\
&\hspace{7cm}-2\Re\left(h_{1}\overline{h_{2}}+h_{3}\overline{h_{4}}\right)\Biggr\}.
\end{split}
\end{equation*}
Here, $W_{5}(\varphi_{1},\varphi_{2},\theta)$ and $W_{6}(\varphi_{1},\varphi_{2},\theta)$ are defined by
\[
W_{5}(\varphi_{1},\varphi_{2},\theta)=|h_{1}|^{2}+|h_{3}|^{2},\qquad 
W_{6}(\varphi_{1},\varphi_{2},\theta)=|h_{2}|^{2}+|h_{4}|^{2}.
\]
 Let $r_{+}(\theta)\equiv|\Lambda_{+}|^{2}$ and $r_{-}(\theta)\equiv|\Lambda_{-}|^{2}$. Remark that
\[
r_{+}(\theta)>1,\qquad 0<r_{-}(\theta)<1.
\]
We put $\Lambda_{+}=r_{1}e^{i\theta_{1}}$ and $\Lambda_{-}=r_{2}e^{i\theta_{2}}$. Since $\Lambda_{+}\cdot \overline{\Lambda_{-}}=1$, we get
\[
r_{2}=\frac{1}{r_{1}},\qquad\theta_{1}=\theta_{2}+2n\pi\qquad(n\in\mathbb{Z}).
\]
Then we obtain
\begin{equation*}
\begin{split}
\mu(x)&=\left|\frac{1}{\Lambda_{+}-\Lambda_{-}}\right|^{2}\Biggl\{(r_{+}(\theta)^{x}(|h_{1}|^{2}+|h_{3}|^{2})+r_{-}(\theta)^{x}(|h_{2}|^{2}+|h_{4}|^{2})\\
&\hspace{7cm}-2\Re\left(h_{1}\overline{h_{2}}+h_{3}\overline{h_{4}}\right)\Biggr\}\\
&=\left|\frac{1}{\Lambda_{+}-\Lambda_{-}}\right|^{2}\Biggl\{(r_{+}(\theta)^{x}(|h_{1}|^{2}+|h_{3}|^{2})+\left(\frac{1}{r_{+}(\theta)}\right)^{x}(|h_{2}|^{2}+|h_{4}|^{2})\\
&\hspace{7cm}-2\Re\left(h_{1}\overline{h_{2}}+h_{3}\overline{h_{4}}\right)\Biggr\}.\\
\end{split}
\end{equation*}
Furthermore, we denote
\[
\Gamma_{+}=-\Lambda_{+},\qquad\Gamma_{-}=-\Lambda_{-}.
\]
For $x\leq-1$, we get
\begin{equation*}
\begin{split}
\mu(x)&=\left|\frac{1}{\Lambda_{+}-\Lambda_{-}}\right|^{2}\Biggl\{(r_{+}(\theta)^{-x}(|k_{1}|^{2}+|k_{3}|^{2})+r_{-}(\theta)^{-x}(|k_{2}|^{2}+|k_{4}|^{2})\\
&\hspace{7cm}-2\Re\left(k_{1}\overline{k_{2}}+k_{3}\overline{k_{4}}\right)\Biggr\}\\
&=\left|\frac{1}{\Lambda_{+}-\Lambda_{-}}\right|^{2}\Biggl\{(r_{+}(\theta)^{-x}(|k_{1}|^{2}+|k_{3}|^{2})+\left(\frac{1}{r_{+}(\theta)}\right)^{-x}(|k_{2}|^{2}+|k_{4}|^{2})\\
&\hspace{7cm}-2\Re\left(k_{1}\overline{k_{2}}+k_{3}\overline{k_{4}}\right)\Biggr\}.\\
\end{split}
\end{equation*}
Here $k_{1}$ and $k_{2}$ are given by
\begin{equation*}
k_{1}=\Psi^{L}(-1)-\Gamma_{-}\varphi_{1},\qquad k_{2}=\Psi^{L}(-1)-\Gamma_{+}\varphi_{1}
\end{equation*}
and $k_{3}$ and $k_{4}$ are given by
\begin{equation*}
k_{3}=\Psi^{R}(-1)-\Gamma_{-}\varphi_{2},\qquad k_{4}=\Psi^{R}(-1)-\Gamma_{+}\varphi_{2}.
\end{equation*}
Since $\sum_{j=1}^{2}|\ell_{j}|^{2}\ne0$ and $\sum_{j=3}^{4}|\ell_{j}|^{2}\ne0\ (\ell=h,k)$ , we have
\[
\mu\in\mathcal{M}_{s,exp}(U_{H}).
\]
\hspace{\fill} $\square$
\vspace{0.3cm}

\subsection{Proof of Theorem $2$}
We put $\widetilde{\mathcal{M}}$ as
\[
\widetilde{\mathcal{M}}\equiv\widetilde{\mathcal{M}_{s,qp}(U_{H})}\cup\widetilde{\mathcal{M}_{s,bdd}(U_{H})}\cup\widetilde{\mathcal{M}_{s,exp}(U_{H})}.
\]
At first, we show that $\widetilde{\mathcal{M}}\subset\widetilde{\mathcal{M}_{s}(U_{H})}$ . This statement is trivial by the definition. Let us show that $\widetilde{\mathcal{M}_{s}(U_{H})}\subset\widetilde{\mathcal{M}}$. For any $\mu\in\widetilde{\mathcal{M}_{s}(U_{H})}$, there exists $\lambda\in S^{1}$ such that $\mu\in\mathcal{M}_{s}^{(\lambda)}(U_{H})$. We put $\lambda=e^{i\theta}$, where $\theta\in K=K_{1}\cup K_{2}\cup K_{3}$. By using Proposition \ref{prop1}, \ref{prop2}, and \ref{prop3}, we obtain
\[
\begin{cases}
\theta\in K_{1}\Longrightarrow\mu\in\widetilde{\mathcal{M}_{s,qp}(U_{H})}\ \mbox{or}\ \mu\in\mathcal{M}_{unif}(U_{H})\\
\theta\in K_{2}\Longrightarrow\mu\in\widetilde{\mathcal{M}_{s,bdd}(U_{H})}\\
\theta\in K_{3}\Longrightarrow\mu\in\widetilde{\mathcal{M}_{s,exp}(U_{H})}
\end{cases}.
\]
From this, the theorem follows.
\hspace{\fill} $\square$
\vspace{0.3cm}

\noindent{\bf{Remark\ 5.}} We consider the spectrum $\sigma(U_{H})$ of the time evolution operator $U_{H}$ for the Hadamard walk on $\mathbb{Z}$. Grimmett et al. \cite{gjs} have derived a weak limit theorem for the quantum walk on $\mathbb{Z}$ based on the Fourier transform. This method (the GJS method) is useful to obtain the spectrum $\sigma(U_{H})$. Now, we briefly see the GJS method and refer the interested readers to \cite{gjs}. Let $f:[-\pi,\pi)\longrightarrow\mathbb{C}^{2}$ and $k\in[-\pi,\pi)$. The Fourier transform of the function $f$ is defined by the integral
\begin{align*}
(\mathcal{F}f)(x)=\frac{1}{2\pi}\int_{-\pi}^{\pi}e^{ikx}f(k)\ dk\qquad(x\in\mathbb{Z}).
\end{align*}
Then the inverse of the Fourier transform $\mathcal{F}^*$ is given by 
\begin{align*}
\hat{g}(k)\equiv(\mathcal{F}^*g)(k)=\sum_{x\in\mathbb{Z}}e^{-ikx}\ g(x)\quad\left(g:\mathbb{Z}\longrightarrow\mathbb{C}^{2},\ k\in [-\pi,\pi)\right).
\end{align*}
From the inverse of the Fourier transform and Eq. \eqref{timeevo}, we have
\begin{align*}
\hat{\Psi}_{n+1}(k)=\hat{U}_C(k)\hat{\Psi}_n(k),
\end{align*}
where $\Psi_n:\mathbb{Z}\longrightarrow\mathbb{C}^{2}$ and matrix $\hat{U}_C(k)$ is determined by
\begin{align*}                          
\hat{U}_C(k)=e^{ik}\begin{bmatrix}
1&0\\
0&0\\
\end{bmatrix}C+e^{-ik}\begin{bmatrix}
0&0\\
0&1\\
\end{bmatrix}C. 
\end{align*}
We remark that matrix $\hat{U}_C(k)$ is a unitary matrix. If we take the Hadamard coin, we have
\[
\hat{U}_H(k)=\begin{bmatrix}
\frac{1}{\sqrt{2}}e^{ik}&\frac{1}{\sqrt{2}}e^{ik}\\
\frac{1}{\sqrt{2}}e^{-ik}&-\frac{1}{\sqrt{2}}e^{-ik}\\
\end{bmatrix}
.
\] 
Thus, the eigenvalues of $\hat{U}_H(k)$ are given by
\[
\lambda_{1}(k)=\frac{\sqrt{1+\cos^{2}k}+i\sin k}{\sqrt{2}},\qquad
\lambda_{2}(k)=\frac{-\sqrt{1+\cos^{2}k}+i\sin k}{\sqrt{2}}.
\]
From the above argument, we get
\[
\sigma(U_{H})=\{e^{i\xi}:\xi\in K_{1}\cup K_{2}\}=\widetilde{K_{1}}\cup\widetilde{K_{2}}.
\]
Here the definitions of $K_{1}$ and $K_{2}$ are given in Sec. 5.1. In terms of the spectral analysis, Morioka \cite{hm} showed that the generalized eigenfunctions are not square summable but belong to $\ell^{\infty}$-space on $\mathbb{Z}$. Namely, the generalized eigenfunction $\Psi$ satisfied with $U_{H}\Psi=\lambda\Psi$ belongs to $\ell^{\infty}$-space, where $\lambda\in\sigma(U_{H})\setminus\widetilde{K_{1}}$.
\section{Summary}\label{Summary}
The present paper dealt with stationary measures of the Hadamard walk on $\mathbb{Z}$. By solving the eigenvalue problem via the transfer matrices $T^{+}_{\lambda}(H)$ and $T^{-}_{\lambda}(H)$, all the stationary measures $\widetilde{\mathcal{M}_{s}(U_{H})}$ were divided into three classes, i.e., quadratic polynomial type $\widetilde{\mathcal{M}_{s,qp}(U_{H})}$, bounded type $\widetilde{\mathcal{M}_{s,bdd}(U_{H})}$, and exponential type $\widetilde{\mathcal{M}_{s,exp}(U_{H})}$. In other words, we obtained 
\[
\widetilde{\mathcal{M}_{s}(U_{H})}=\widetilde{\mathcal{M}_{s,qp}(U_{H})}\cup\widetilde{\mathcal{M}_{s,bdd}(U_{H})}\cup\widetilde{\mathcal{M}_{s,exp}(U_{H})}.
\]
In particular, we presented an explicit necessary and sufficient condition for the bounded-type stationary measure to be periodic. Furthermore, we confirmed that any stationary measure in $\widetilde{\mathcal{M}_{s}(U_{H})}$ is not probability measure. This result is strikingly different from the corresponding one for three-state Grover walk on $\mathbb{Z}$. In fact, the set of stationary measures for this walk contains $\ell^{2}$-function and functions with finite support. It would be an interesting future problem to prove that 
\[
\mathcal{M}_{s}(U_{H})=\widetilde{\mathcal{M}_{s}(U_{H})}.
\]
\par
\noindent \\
\noindent \\\noindent \\
\section*{Conflict of interest}
On behalf of all authors, the corresponding author states that there is no conflict of interest.


\begin{thebibliography}{10}
\bibitem{adz}
Y. Aharonov, L. Davidovich and N. Zagury, {\it Quantum random walks}, Phys. Rev. A \textbf{48}, pp.1687-1690 (1993).
\bibitem{Ahl2011}
A. Ahlbrecht, V. B. Scholz and A. H. Werner, \textit{Disordered quantum walks in one lattice dimension}, J. Math. Phys. {\bf 52}, 102201 (2011).
\bibitem{akr15}
A. Ambainis, J. Kempe and A. Rivosh, {\it Coins make quantum walks faster}, Proceedings of the Sixteenth Annual ACM-SIAM Symposium on Discrete Algorithms, SIAM, Philadelphia, pp.1099-1108 (2005).

\bibitem{Bou2003} 
O. Bourget, J. S. Howland and A. Joye, \textit{Spectral analysis of unitary band matrices}, Comm. Math. Phys. {\bf 234}, 191-227 (2003).



\bibitem{gjs}
G. Grimmett, S. Janson and P. F. Scudo, {\it Weak limits for quantum random walks}, Phys. Rev. E \textbf{69}, 026119 (2004).
\bibitem{Kawai2017}
H. Kawai, T. Komatsu and N. Konno,
\textit{Stationary measures of three-state quantum walks on the one-dimensional lattice},
 Yokohama Math. J. \textbf{63}, 59-74 (2017). 
 \bibitem{Kawai2018}
H. Kawai, T. Komatsu and N. Konno,
\textit{Stationary measure for two-state space-inhomogeneous quantum walk in one dimension}, Yokohama Math. J. {\bf{64}}, 111-130 (2018).

\bibitem{KKMS1}
T. Komatsu, N. Konno, H. Morioka and E. Segawa, 
\textit{Generalized eigenfunctions for the quantum walks via a path counting approach}, Reviews in Mathematical Physics \textbf{33}, 2150019 (2021).
\bibitem{Komatsu2017}
T. Komatsu and N. Konno, 
\textit{Stationary amplitudes of quantum walks on the higher-dimensional integer lattice}, Quantum Inf. Process. {\bf16}, 291 (2017).




\bibitem{ko1}
N. Konno, {\it A new type of limit theorems for the one-dimensional quantum random walk}, J. Math. Soc. Japan \textbf{57}, pp.1179-1195 (2005).
\bibitem{ko2}
N. Konno, {\it The uniform measure for discrete-time quantum walks in one dimension}. Quantum Inf. Process.  \textbf{13}, pp.1103-1125 (2014).
\bibitem{kls}
N. Konno, T. Luczak and E. Segawa, {\it Limit measures of inhomogeneous discrete-time quantum walks in one dimension}, Quantum Inf. Process. \textbf{12}, pp.33-53 (2013).
\bibitem{kt}
N. Konno and M. Takei, {\it The non-uniform stationary measure for discrete-time quantum walks in one dimension}, Quantum Inf. Comput. \textbf{15}, pp.1060-1075 (2015).
\bibitem{mw}
K. Manouchehri and J. Wang, {\it Physical Implementation of Quantum Walks}, Springer (2013).
\bibitem{MMOS}
K. Matsue, L. Matsuoka, O. Ogurisu and E. Segawa, 
\textit{Resonant-tunneling in discrete-time quantum walk},
Quantum Studies: Mathematics and Foundations {\bf 6}, 35-44 (2018). 
\bibitem{M1}
A. Messiah, \textit{Quantum Mechanics Volume $1$}, North-Holland, Amsterdam (1961).
\bibitem{hm}
H. Morioka, {\it Generalized eigenfunctions and scattering matrices for position-dependent quantum walks}, Rev. Math. Phys., (2019).
\bibitem{s1}
A. Suzuki, {\it Asymptotic velocity of a position-dependent quantum walk}, Quantum Inf. Process. \textbf{15}, 103-119 (2016).
\bibitem{por}
R. Portugal, {\it Quantum Walks and Search Algorithms}, Springer (2013).
\end{thebibliography}
\end{document}